\documentclass[a4paper,11pt]{article}
\pdfoutput=1

\usepackage{jheppub}
\usepackage{graphicx,epsfig}
\usepackage{simplewick}
\usepackage{slashed}
\usepackage{color}
\usepackage{setspace}
\usepackage{hyperref}
\usepackage{xspace}

\def\be{\begin{equation}}
\def\ee{\end{equation}}
\def\bea{\begin{eqnarray}}
\def\eea{\end{eqnarray}}
\def\bear{\begin{array}}
\def\eear{\end{array}}
\def\bes{\begin{subequations}}
\def\ees{\end{subequations}}

\def\lQ{\Lambda_{\rm QCD}}

\newcommand{\MSbar}{\overline{\rm MS}}  
\newcommand{\m}{{\overline m}}

\usepackage{graphics}
\usepackage{graphicx}
\usepackage{dcolumn}
\usepackage{bm}
\usepackage{epsfig}
\usepackage{graphicx}
\usepackage{multirow}
\usepackage{dcolumn}
\usepackage{graphicx,epsfig}%

\newcommand{\blue}{\textcolor{blue}}

\newcommand{\MS}{\overline{\rm MS}}
\newcommand{\RS}{\rm RS}

\newcommand{\nn}{\nonumber}
\def\als{\alpha_{s}}

\newcommand{\eq}[1]{Eq.~\eqref{#1}}
\newcommand{\eqs}[2]{Eqs.~\eqref{#1} and \eqref{#2}}

\newcommand{\fig}[1]{Fig.~\ref{#1}}
\newcommand{\Sec}[1]{Sec.~\ref{#1}}
\newcommand{\Tab}[1]{Table~\ref{#1}}

\newcommand{\rcite}[1]{Ref.~\cite{#1}}
\newcommand{\rcites}[1]{Refs.~\cite{#1}}

\title{The charm/bottom quark mass from heavy quarkonium at N$^3$LO}

\author[a]{Clara Peset}
\author[b]{Antonio Pineda}
\author[b]{Jorge Segovia}

\emailAdd{clara.peset@tum.de}
\emailAdd{pineda@ifae.es}
\emailAdd{jsegovia@ifae.es}

\affiliation[a]{Physik Department T31, James-Franck-Stra\ss e 1, Technische Universit\"at M\"unchen,
85748 Garching, Germany}
\affiliation[b]{Grup de F\'\i sica Te\`orica, Dept. F\'\i sica and IFAE-BIST, Universitat Aut\`onoma de Barcelona,\\ 
E-08193 Bellaterra (Barcelona), Spain}

\date{\today}

\abstract{
We determine the charm and bottom quark masses using the N$^3$LO perturbative expression of the ground state 
(pseudoscalar) energy of the bottomonium, charmonium and $B_c$ systems:
the $\eta_b$, $\eta_c$ and $B_c$ masses. 
We work in the renormalon subtracted scheme, which allows us to control the divergence of the
 perturbation series due to the pole mass renormalon. 
Our result for the $\MSbar$ masses reads $\m_{c}(\m_{c})=1223(33)$ MeV and $\m_{b}(\m_{b})=4186(37)$ MeV. We also extract a value of $\als$ from a renormalon-free combination of the $\eta_b$, $\eta_c$ and $B_c$ masses: $\als(M_z)=0.1195(53)$. We explore the applicability of the weak coupling approximation to bottomonium $n=2$ states. Finally, we consider an alternative computational scheme that treats the static potential exactly and study its convergence properties.}

\begin{document}

\preprint{
\begin{flushright}
TUM-HEP-1144/18
\end{flushright}
}

\maketitle


\section{Introduction}
\label{intr}

The bottomonium and charmonium (heavy quarkonium systems where the two nonrelativistic constituents have equal masses) masses
have been computed to increasingly higher order in perturbation theory over the years \cite{Billoire:1979ih,Pineda:1997hz,BPSV,Kniehl:2002br,Penin:2002zv,Penin:2005eu,Beneke:2005hg,Kiyo:2013aea,Kiyo:2014uca}, presently reaching next-to-next-to-next-to-leading order (N$^3$LO) precision, i.e. ${\cal O}(m\als^5)$. The use of effective field theories was instrumental in getting this precision 
\cite{Caswell:1985ui,Bodwin:1994jh, Pineda:1997bj,Brambilla:1999xf}.

The cancellation of the leading renormalon of the pole mass and the static potential, first found in \cite{Pineda:1998id}, and later in \cite{Hoang:1998nz,Beneke:1998rk}, led to the realization \cite{Beneke:1998rk} that using threshold masses \cite{Bigi:1994em,Beneke:1998rk,Pineda:2001zq,Hoang:2009yr,Brambilla:2017hcq} (which explicitly implement the cancellation of the renormalon in heavy quarkonium observables) improves the convergence of the perturbative series. This has made these very precise computations useful not only for academical purposes but also for phenomenological applications. As a consequence, determinations of the bottom, or bottom and charm quark masses have been obtained using these new results
\cite{Beneke:1999fe,Brambilla:2001fw,Pineda:2001zq,Brambilla:2001qk,Lee:2003hh,Ayala:2014yxa,Kiyo:2015ufa,Ayala:2016sdn,Mateu:2017hlz}. 

More recently, the $B_c$ spectrum has also reached N$^3$LO precision \cite{Peset:2015vvi}. This theoretical expression has not yet been used for phenomenology and confronted with experiment.  One can use the $B_c$ spectrum to determine the heavy quark masses. In principle this system is more perturbative than charmonium. Hence, it should lead to a more accurate determination of the charm quark mass. We aim to do so in this paper. In particular, we consider specific energy combinations that are more suitable for clean theoretical analyses. This results in more accurate determinations of the charm quark mass and also allows an independent determination of $\als$.

The other main motivation of this paper is to study the feasibility of an alternative computational scheme that reorganizes the perturbative expansion of the above analyses. This scheme is characterized by solving the Schroedinger equation including the static potential exactly (to the order it is known). This incorporates formally subleading terms in the leading order (LO) solution. 
On the other hand the relativistic corrections to the spectrum are included perturbatively. This working scheme performs a partial resummation of higher order effects. This may accelerate the convergence of the perturbative series. This is indeed the effect seen in the cases where it has been applied (spectrum and decays) \cite{Recksiegel:2002za,Recksiegel:2003fm,Kiyo:2010jm,Pineda:2013lta}. This scheme naturally leads to the organization of the computation in powers of $v$, the relative velocity of the heavy quark in the bound state.

Irrespectively if working with strict fixed-order perturbation theory or with an improved perturbation scheme, one has to implement the renormalon cancellation in the computation. In this paper we use the RS threshold mass as the expansion parameter \cite{Pineda:2001zq}. 

Finally, we explore the applicability of the weak-coupling approximation to the first excited state ($n=2$) of heavy quarkonium, both in the strict weak-coupling approximation and in the alternative perturbation scheme. For this analysis we mainly consider renormalon-free combinations, which are theoretically cleaner.

\section{Ground state Heavy Quarkonium energy at N$^3$LO}
\label{Sec:mbot}

In this section we determine the bottom and charm quark masses from the experimental values 
(as quoted from \rcite{Patrignani:2016xqp}) of the ground state masses of the bottomonium, charmonium and $B_c$, and the corresponding theoretical expressions to N$^3$LO. Whereas the bottomonium and charmonium spectra have been used before, the use of the $B_c$ spectrum is novel.

An analysis of the $\Upsilon(1S)$ mass to N$^3$LO in the strict weak-coupling expansion, and using the RS threshold mass, has been done in \rcites{Ayala:2014yxa,Ayala:2016sdn}. Here we will follow a similar methodology but considering instead the $\eta_b$.\footnote{Here and throughout this paper we refer to $\eta_{b,c}(1S)$ as $\eta_{b,c}$, $B_c(1S)$ as $B_c$ and $\eta_{b,c}(2S)$ as $\eta'_{b,c}$.} The underlying reason for using pseudoscalar masses in this paper is that the experimental information of the $B_c$ spectrum is incomplete. For the ground state, only the pseudoscalar mass is known. 
Therefore, the experimental set of masses we will use are the $\eta_b$, $\eta_c$ and $B_c$ 
masses. 
 
\subsection{Determination of $m_b$}
\label{Subsec:mb}
We extract the bottom quark mass from the condition:
\begin{align}
M^{(th)}_{\eta_b}
&= M^{(exp)}_{\eta_b}= 9.3990(23)\ {\rm GeV}\label{condetab}
\,.\end{align}

The determination of the bottom quark mass and the error analysis closely follows \rcites{Ayala:2014yxa,Ayala:2016sdn} (see those references for extra details and notation). We vary the parameters in the following way: $\nu_f=2^{+1}_{-1}$ GeV, $\als(M_z)=0.1184(12)$ \cite{Patrignani:2016xqp} (with decoupling at
$\m_b= 4.2$ GeV and at $\m_c=1.27$ GeV; the specific location of the decoupling plays a marginal role in the determination of the bottom quark mass), $N_m=0.5626(260)$, 
and $(4/3) r_3(\m_b;N_l)=1698.59 \pm 1.74$ \cite{Marquard:2016dcn}.\footnote{Note that there have been some small changes in 
$\als$ and $r_3$ with respect to \rcites{Ayala:2014yxa,Ayala:2016sdn}. Indeed with the new value of $r_3$ (the four loop coefficient relating the pole and the $\MS$ mass) the associated error is negligible: $\sim 0$ MeV in all cases. Therefore, we will not explicitly display it in the following.} We determine the central value (and the renormalization scale associated error) using $\nu=5^{+3}_{-3}$ GeV. This scale is bigger than the scale used in \rcites{Ayala:2014yxa,Ayala:2016sdn},  which was $\nu=2.5_{-1}^{+1.5}$GeV. This is motivated by inspecting the scale dependence of the observable. In the RS scheme, 
$\nu \sim 5$ GeV is roughly the scale where the function is extremal. The RS' scheme suggests even higher scales. For scales smaller than 2 GeV the result shows a strong scale dependence. This motivates the minimum that we take for the $\nu$, and we take the maximum symmetrically, thus fixing the scale variation. 

Using \eq{condetab} in the RS and RS' approaches we extract, in MeV, respectively
\bea
\label{MbRSdet1}
m_{b,\RS}(2\;{\rm GeV})
&=&
4\, 379_{+31}^{+1}(\nu)^{-4}_{+5}(\nu_f)^{-5}_{+5}(\als)^{-32}_{+32}(N_m);
\\
\label{MbMSRSdet1}
\Rightarrow 
\;\;\;\;   \m_{b}
&=&{4\,185}_{+28}^{+1}(\nu)^{-4}_{+5}(\nu_f)^{-10}_{+10}(\als)^{+8}_{-8}(N_m).
\\
\label{MbRSpdet1}
m_{b,\RS'}(2\;{\rm GeV})&=&4\,742^{-10}_{+39}(\nu)^{-2}_{+3}(\nu_f)
^{+4}_{-4}(\als)^{-15}_{+15}(N_m);
\\
\label{MbMSRSpdet1}
\Rightarrow 
\;\;\;\;   \m_{b}
&=&4\,183^{-9}_{+35}(\nu)^{-2}_{+3}(\nu_f)^{-10}_{+10}(\als)^{+8}_{-8}(N_m).
\eea
In \eqs{MbMSRSdet1}{MbMSRSpdet1}, $\m_b$ has been determined from the RS masses at the scale $\nu=2.5$ GeV. Setting $\nu=5$ GeV would not change much the value: bigger scales produce small changes in the value of the mass but smaller scales produce larger variations. In both cases reasonable renormalization scale variations in the relation between the RS masses and $\m_b$ are well inside the uncertainties of our determination of $\m_b$. Overall, the uncertainties in $\m_b$ are dominated by the variation of the renormalization scale $\nu$ in the fit. 

\begin{figure}[htb]
	\begin{center}      
	\includegraphics[width=0.48\textwidth]{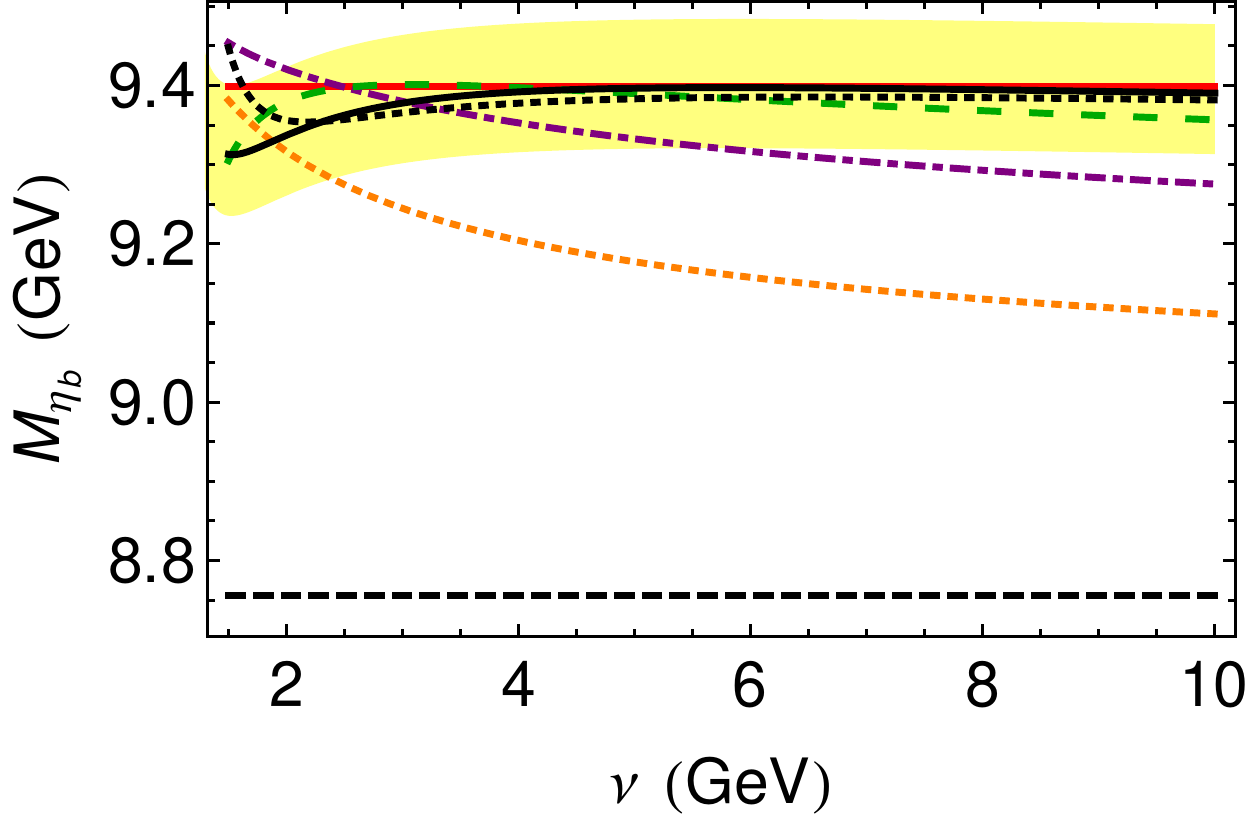}
	%
		\includegraphics[width=0.51\textwidth]{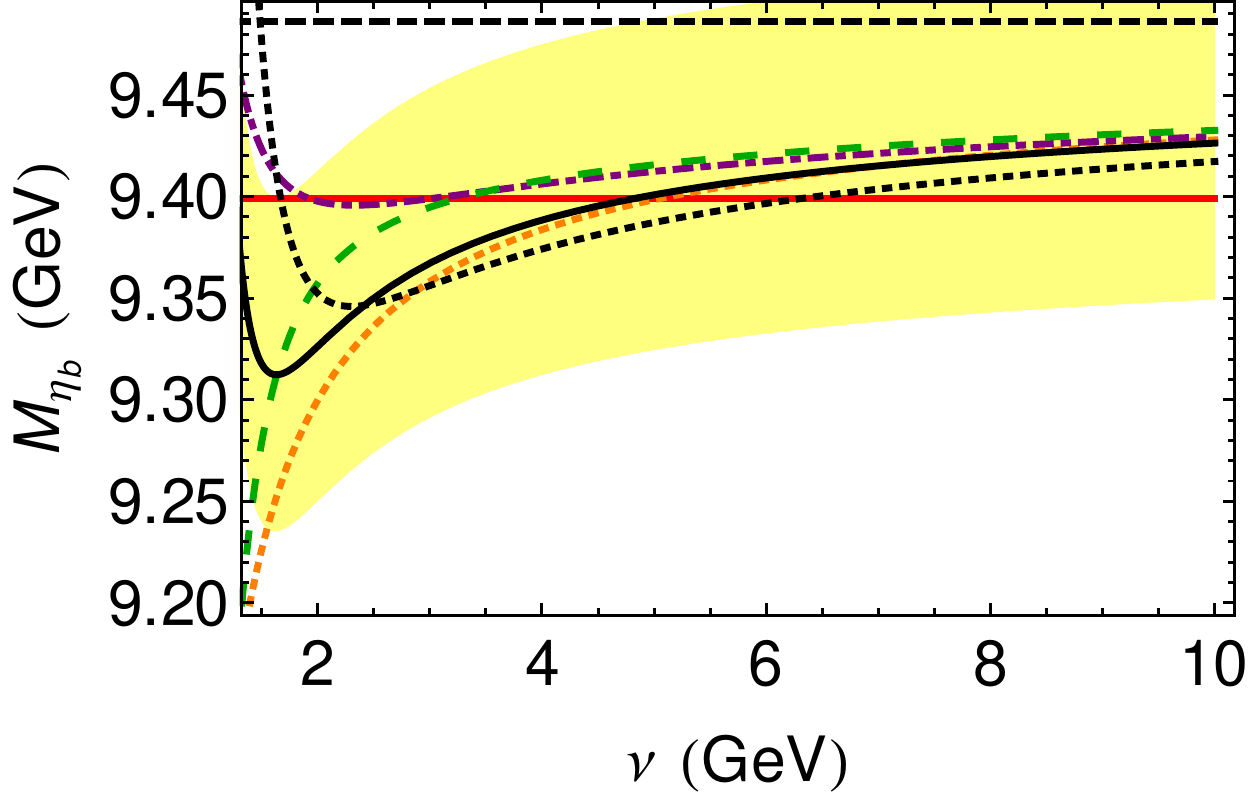}
\caption{Plots of $M_{\eta_b}$ for $\nu_f=2$ GeV, $m_{b,\RS}=4.378^{-41}_{+41}(\m_b)$ GeV, $m_{b,\RS'}=4.743^{-41}_{+41}(\m_b)$ GeV. Red line is the experimental value, black dashed line is $2m_{b,\RS/\RS'}$. Orange dotted, purple dot-dashed and green long-dashed are the LO-NNLO contributions respectively. Solid black line is the N$^3$LO, and the dotted black line the N$^3$LO without $\delta E_{10}^{US}$. The yellow band corresponds to the error of
$m_{b,\RS/\RS'}$ associated to the uncertainty of $\m_b$.
 {\bf Left panel:} Plot of $M_{\eta_b}$ in RS scheme.
{\bf Right panel:} Plot of $M_{\eta_b}$ in RS' scheme.
\label{Fig:etab} }  
\end{center}
\end{figure}

Taking the average of \eqs{MbMSRSdet1}{MbMSRSpdet1} we obtain
\be
\m_b  =  4184(37) \; {\rm MeV} \ ,
\label{resultmb}
\ee 
where we have rounded the $\pm$ variation of each parameter to the maximum (for the combined error we take the largest of both determinations) and added them in quadrature. To this number we add the leading charm corrections following the analysis of \cite{Ayala:2014yxa}. It is summarized by a shift in the final value of $\sim +2$ MeV. Our final number reads
\be
\m_b  =  4186(37)\; {\rm MeV} \;.
\label{resultmbfinal}
\ee 
For illustration, we show the value of the different orders in the perturbation series for the central value parameters:
\bea
 {\rm RS: \;} M_{\eta_b}&=& 
 ( 
8759 + 421 + 155 + 57 + 7
 ) \ {\rm MeV} \,, 
\\
{\rm RS': \;} M_{\eta_b}&=&
 ( 
9484 - 88 + 14 + 3 - 14
 ) \ {\rm MeV} \,, 
\\
m_{b, \RS}(2\;{\rm GeV})&=&
 ( 
4185 + 145 + 58 + 9 - 18
 ) \ {\rm MeV} \ ,
\\
m_{b, \RS'}(2\;{\rm GeV})&=&
 ( 
4183 + 473 + 86 + 16 - 16
 ) \ {\rm MeV} \ .
\eea
Overall, we see a convergent pattern. Let us now study in more detail the convergence of the perturbative series, and the reliability of the error estimate in \eq{resultmbfinal}. In \fig{Fig:etab} we study the convergence of the perturbative expansion of $M_{\eta_b}$ and its dependence on the factorization scale $\nu$, which is the major source of uncertainty. We do so for both the RS and RS' schemes (even though the differences between both schemes are large at low orders they become quite small at N$^3$LO). 

\begin{table}[h]
\begin{center}
\begin{tabular}{|c|c|c|c|}
  \cline{1-4}
\rule[-1ex]{0pt}{0pt} \rule{0pt}{3ex} $\nu_f$ &0.7 GeV & 1 GeV & 2 GeV \\
  \hline \hline
\rule[-2.5ex]{0pt}{0pt}\rule{0pt}{4ex}  $m_{b,\RS}$ (MeV)& 4\,717(41) &4\,618(41) & 4\,378(41) \\
 \hline
  \rule[-2.5ex]{0pt}{0pt}\rule{0pt}{4ex}  $m_{b,\RS'}$ (MeV) & 4\,949(41) & 4\,885(41) & 4\,743(41)  \\
  \hline
\end{tabular}
\caption{RS and RS' bottom quark masses for different values of $\nu_f$.}
\label{Tab:mb}
\end{center}
\end{table}
In \Tab{Tab:mb}, we use the central value in \eq{resultmb} to determine the RS masses (for latter use we also give the values for $\nu_f=0.7,\; 1$ GeV).
The difference between the values in the last column of \Tab{Tab:mb} and the values in \eqs{MbRSdet1}{MbRSpdet1} is completely marginal. 
In the values displayed in \Tab{Tab:mb} we explicitly incorporate the error from \eq{resultmb} to show the sensitivity to the error of $\m_b(\m_b)$. As expected, and needed for a precise determination of the heavy quark mass, we find such sensitivity to be very large (see the yellow bands in \fig{Fig:etab}). 

In general, we observe that the convergence is quite reasonable irrespectively of the scheme (with $\nu_f=1$ GeV the RS' scheme does not converge that well, though the final value of the bottom quark mass is very similar); except for the last (N$^3$LO) term, where the factorization scale dependence becomes stronger. This could signal the importance of ultrasoft effects. In order to quantify their importance we have also plotted the N$^3$LO prediction of $M_{\eta_b}$ minus $\delta E^{US}_{10}$, the pure ultrasoft contribution to the energy, which for a general quantum number reads
\begin{align}
\label{EnlUS}
\delta E_{nl}^{US}(\nu)&=\frac{m_r C_F^2\als^2}{2n^2}
\frac{\als^3(\nu)}{\pi}\left[\frac{2}{3} C_F^3 L^E_{nl}+\frac{1}{3} C_A\left(
\ln\frac{\nu n^2}{m_r C^2_F\als^2}
+\frac{5}{6}\right) \left(\frac{C_A^2}{2}+\frac{4 C_A C_F}{(2 l+1) n}\right.\right.\nn\\
&\quad +\left.\left.2C_F^2 \left(\frac{8}{(2 l+1) n}-\frac{1}{n^2}\right)\right)+\frac{8\delta_{l0} }{3 n}C_F^2  \left(C_F-\frac{C_A}{2}\right) \left(\ln\frac{\nu n^2}{m_r C^2_F\als^2}+\frac{5}{6}\right)\right],
\end{align}
where $m_r=m_b/2$ for bottomonium, $m_r=m_c/2$ for charmonium, and $m_r=m_bm_c/(m_b+m_c)$ for the $B_c$. $L^E_n$
are the non-Abelian Bethe logarithms, numerically determined in \rcite{Kniehl:1999ud} for $l=0$ and in \rcite{Kiyo:2014uca} for $l\not=0$ for low values of $n$. 

We observe that for scales larger than 2 GeV, $\delta E^{US}_{10}$ is small and well inside uncertainties. It is also comforting that the yellow error band in \fig{Fig:etab} (associated to the uncertainty of $\m_b$) covers the difference between the NNLO and N$^3$LO results in the RS and in the RS' scheme for the whole scale variation. This difference is often used as an alternative determination of the error. Actually, in the RS' scheme the distance between the different orders is quite small, which may ask for future more aggressive determinations of the bottom mass. We refrain from doing so until a more clear understanding of the renormalization scale dependence of the observable at low scales and of the role played by the ultrasoft scale is achieved. 

As a final check, we have performed the fit to the $\Upsilon(1S)$ mass using the same setup. We find a shift of almost +20 MeV, well within the uncertainties of \eq{resultmb}.

\begin{figure}[!htb]
	\begin{center}   
	\includegraphics[width=0.85
\textwidth]{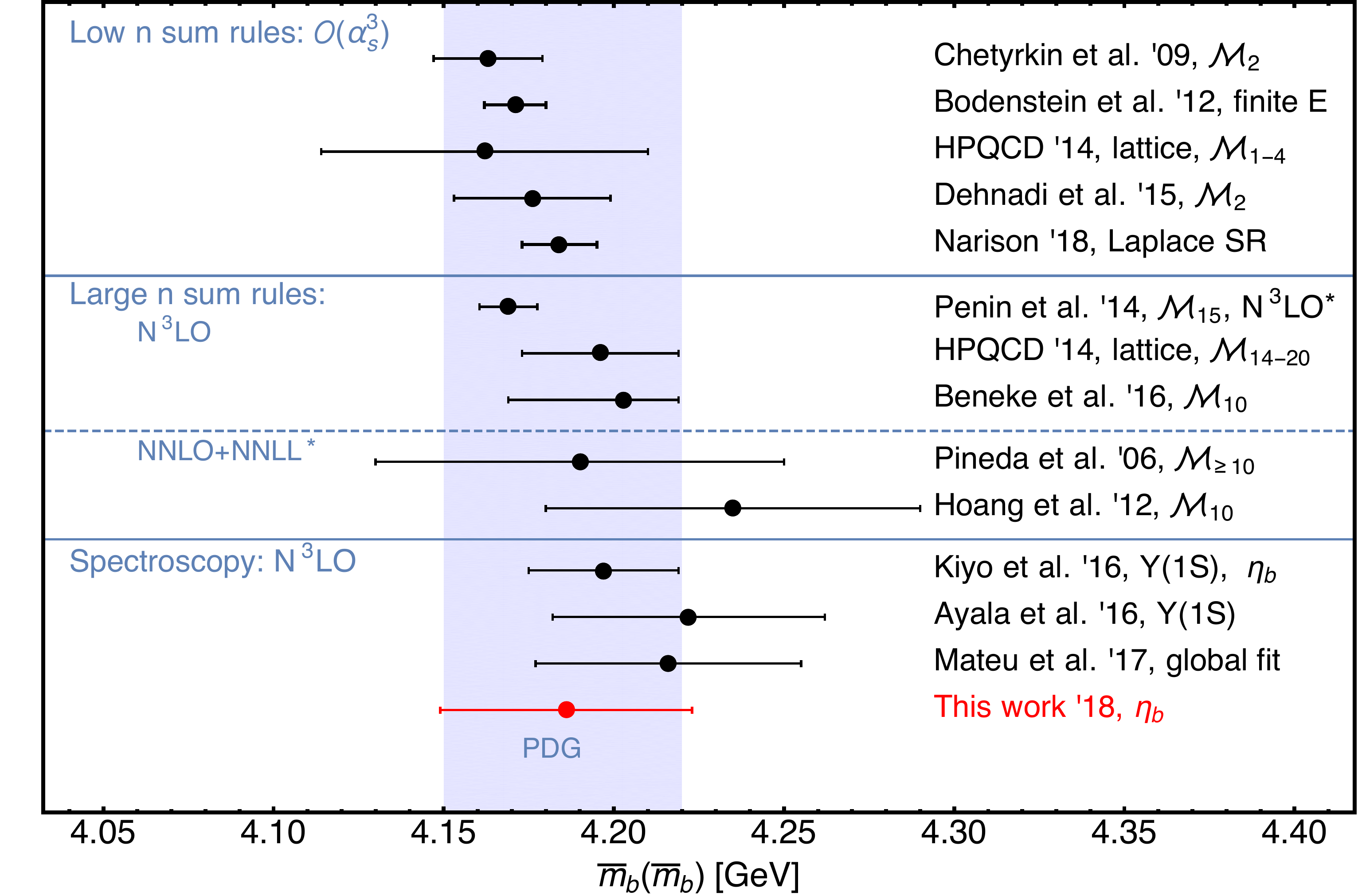}	      
\caption{$\m_{b}(\m_{b})$ determined from spectroscopy or from sum rules in the strict weak-coupling approximation. 
Chetyrkin et al. \cite{Chetyrkin:2009fv}, HPQCD14 \cite{Chakraborty:2014aca}, and Dehnadi et al. \cite{Dehnadi:2015fra} 
use N$^3$LO low n moments equated to the corresponding experimental moments evaluated using the available experimental information (masses and decays) supplemented with lattice information and/or assuming perturbation theory at high energies. Finite energy \cite{Bodenstein:2011fv} or Laplace \cite{Narison:2018dcr} sum rules have also been applied for low $n$. 
Penin et al. \cite{Penin:2014zaa} and Beneke et al. \cite{Beneke:2014pta,Beneke:2016oox} uses nonrelativistic (with Coulomb resummation) N$^3$LO large $n$ moments (the first reference uses a partial result). 
Pineda et al. \cite{Pineda:2006gx} and Hoang et al. \cite{Hoang:2012us} use NNLO and (a partial) NNLL expression for nonrelativistic (with Coulomb resummation) large $n$ moments. 
Kiyo et al. \cite{Kiyo:2015ufa}, Ayala et al. \cite{Ayala:2016sdn}, Mateu et al. \cite{Mateu:2017hlz} and this paper use N$^3$LO expressions for the heavy quarkonium masses. \label{Tabmb}} 
\end{center}
\end{figure}

We now compare our result with previous determinations of $\m_{b}(\m_{b})$ from spectroscopy or from sum rules. We show such comparison in Fig. \ref{Tabmb}. We find that all determinations are perfectly consistent with each other. Determinations from sum rules with small number of momentums (low $n$ sum rules) seem to give slightly smaller values of the mass than large $n$ or spectroscopy determinations. We notice that the latter include the Coulomb resummation. Therefore, the physics they are sensitive to is different. Nevertheless, at this stage this difference is not statistically significant, and we are not in the position to 
 push this discussion further. It is also worth noting that, at present, partial NNLL sum rules determinations produce slightly larger errors than other determinations. It would be interesting to see if the complete result, or its combination with the large $n$ N$^3$LO sum result, may lead to a more precise determination. 
 
 We now turn our discussion to the comparison between different spectroscopy determinations. Our value is slightlty smaller than the determinations in \cite{Kiyo:2015ufa,Ayala:2016sdn,Mateu:2017hlz} (but well within each other's uncertanties). Compared with the determinations of Ayala et al. \cite{Ayala:2016sdn}  
and Mateu et al. \cite{Mateu:2017hlz}, a part of the difference can be traced back to the fact that in the two latter references the $\Upsilon$ plays a predominant role in the fit (which leads to a larger value for the bottom mass than fits to the $\eta_b$ mass). In the first reference because the fit was only made to the $\Upsilon$, in the second reference because, even if both the vector and the pseudoscalar were considered in the fit, their method weights the most the particle that produces the smaller errror, which in their paper is the vector. It is worth mentioning that if one looks at the individual fits in \cite{Mateu:2017hlz} to the $\eta_b$ mass one finds a smaller value of $\m_{b}(\m_{b})$, quite close to ours. 
As at present all these determinations are consistent with each other, there is not a clear reason to prefer one versus the other. As we have already discussed before, the resummation of large logarithms may help to solve this problem. We hope to come back to this issue in the future. The other fact that produces a different central value of $\m_{b}(\m_{b})$ is the choice of the factorization scale $\mu$ to fix the central value. In this paper we have chosen a larger value of the factorization scale than in \cite{Ayala:2016sdn,Mateu:2017hlz}. We somewhat feel that a smaller value of the factorization scale leads to a region where the perturbative series does not behave well enough. In this respect the factorization scales we choose are closer to the factorization scales used in \cite{Kiyo:2015ufa}. Actually, the difference with respect to the value obtained in \cite{Kiyo:2015ufa} can be traced back to the fact that in that reference an average of the determination of the bottom mass from the $\Upsilon$ and the $\eta_b$ is made. Again pure determinations from the $\eta_b$ mass are closer to our numbers. It is also worth mentioning that the error of the $\m_{b}(\m_{b})$ in \cite{Kiyo:2015ufa} from the $\Upsilon$ or $\eta_b$ mass is quite similar (this is the reason why the value obtained in \cite{Kiyo:2015ufa} is more or less the aritmetic average between the values of $\m_{b}(\m_{b})$ obtained from the $\Upsilon$ and the $\eta_b$).

\subsection{Determination of $m_c$}

We first extract the charm quark mass from the charmonium system using
\begin{align}
M^{(th)}_{\eta_c}
&= M^{(exp)}_{\eta_c}= 2.9834(5)\ {\rm GeV}\label{condetac}
\,,
\end{align}
and follow the analysis made in the previous section adapted to this case. The central value determination of $m_{c,\RS/\RS'}$ and the error estimates are computed using the following values for the parameters: $\nu=2.5^{+1.5}_{-1.0}$ GeV,
$\nu_f=1^{+0.5}_{-0.3}$ GeV, $\als(M_z)=0.1184(12)$, $N_m=0.5626(260)$, 
and $(4/3) r_3(\m_c;N_l)=1698.59 \pm 1.74$. The scale $\nu \sim 2.5$ GeV is chosen after  inspecting the 
renormalization scale dependence of the observable, since around this value $M^{(th)}_{\eta_c}$ is extremal in the RS scheme. We also observe that for scales smaller than 1.5 GeV there is a strong scale dependence. This fixes the scale variation we take. We also take a smaller value of $\nu_f$, more natural for charmonium. 

Using \eq{condetac} in RS and RS' approaches we extract, in MeV, respectively
\bea
\label{MRSdet1}
m_{c,\RS}(1{\rm GeV})
&=&
1\, 202^{+15}_{+16}(\nu)^{-15}_{+11}(\nu_f)^{-10}_{+10}(\als)^{-34}_{+34}(N_m);
\\
\label{MMSRSdet1}
\Rightarrow 
\;\;\;\;   \m_{c}
&=&1\,217^{+12}_{+13}(\nu)^{-1}_{-6}(\nu_f)^{-8}_{+8}(\als)^{+10}_{-10}(N_m).
\\
\label{MRSpdet1}
m_{c,\RS'}(1\;{\rm GeV})&=&1\,495^{-11}_{+50}(\nu)^{-9}_{+20}(\nu_f)
^{+4}_{-4}(\als)^{-20}_{+20}(N_m);
\\
\label{MMSRSpdet1}
\Rightarrow 
\;\;\;\;   \m_{c}
&=&1\,222^{-9}_{+40}(\nu)^{-7}_{+16}(\nu_f)^{-7}_{+7}(\als)^{+11}_{-11}(N_m),
\eea
where $\m_c$ is fixed from the RS masses at the scale $\nu=1.5$ GeV (considering $\nu=2.5$ or $\nu=1$ GeV produces small changes in $\m_c$).  Nicely enough, the uncertainty associated to the scheme dependence is quite small, i.e. the values obtained in \eqs{MMSRSdet1}{MMSRSpdet1} are quite close. The uncertainties in $\m_c$ are dominated by the variation of the renormalization scale $\nu$. 

Taking the average of the RS and RS' determinations of $\m_c$ we obtain
\be
\m_c  =  1220(45) \; {\rm MeV} \ ,
\label{result1}
\ee 
where we have rounded the $\pm$ variation of each parameter to the maximum (for the combined error we take the largest of both determinations) and added them in quadrature. 

For illustration, we show the value of the different orders in the perturbation series for the central value parameters: 
\bea
 {\rm RS: \;} M_{\eta_c}&=& 
 ( 
2403 + 291 + 155 + 96 + 38
 ) \ {\rm MeV} \,, 
\\
{\rm RS': \;} M_{\eta_c}&=&
 ( 
2989 - 47 + 26 + 23 - 8
 ) \ {\rm MeV} \,, 
\\
m_{c, \RS}(1\;{\rm GeV})&=&
 ( 
1217 - 35 + 18 + 11 - 9
 ) \ {\rm MeV} \ ,
\\
m_{c, \RS'}(1\;{\rm GeV})&=&
 ( 
1222 + 182 + 62 + 29 - 1
 ) \ {\rm MeV} \ . 
\eea
We see that the perturbative expansions are reasonably convergent. 

\medskip

Before we study in more detail the numbers obtained above and the error analysis, we turn to our alternative determination of the charm mass. We aim to use the experimental value of $M^{(exp)}_{B_c}$. Nevertheless, such observable is strongly dependent on the bottom quark mass. To eliminate such dependence it is better to consider the following observable instead:
\begin{align}
M^{(th)}_{B_c}-M^{(th)}_{\eta_b}/2
& = M^{(exp)}_{B_c}-M^{(exp)}_{\eta_b}/2   = 1.5754(14) \ {\rm GeV} 
\,.
\label{condBc}
\end{align}
We follow the same setup used for bottomonium and charmonium. The central value we take, $\nu=3$ GeV, lays between the values used for $\eta_b$ and $\eta_c$. Following the discussion for bottomonium and charmonium, this scale is of the order of the scale where the observable shows an extremal behavior in the renormalization scale in the RS scheme. 
The central value determination of $m_{c,\RS/\RS'}$ and the error estimates are obtained using the following values for the parameters: $\nu=3^{+1.5}_{-1.0}$ GeV,
$\nu_f=1^{+0.5}_{-0.3}$ GeV, $\als(M_z)=0.1184(12)$, $N_m=0.5626(260)$, and $(4/3) r_3(\m_c;N_l)=1698.59 \pm 1.74$. For the bottom mass we use the RS and RS' masses at $\nu_f=1$ GeV shown in \Tab{Tab:mb} that follow from \eq{resultmbfinal}. Overall, 
using \eq{condBc} in RS and RS' approaches we extract, in MeV, respectively
\bea
\label{MRSdet2}
m_{c,\RS}(1{\rm GeV})
&=&1
\,204^{+27}_{-8}(\nu)^{-26}_{+18}(\nu_f)^{-17}_{+16}(\als)^{-33}_{+33}(N_m)^{-1}_{+1}(\m_b);
\\
\label{MMSRSdet2}
\Rightarrow 
\;\;\;\;   \m_{c}
&=&1\,220^{+21}_{-6}(\nu)^{-7}_{-4}(\nu_f)^{-14}_{+13}(\als)^{+11}_{-11}(N_m)^{-1}_{+1}(\m_b).
\\
\label{MRSprimedet2}
m_{c,\RS'}(1\;{\rm GeV})&=&1\,501^{+1}_{+23}(\nu)^{-14}_{-27}(\nu_f)
^{-2}_{+2}(\als)^{-18}_{+18}(N_m)^{-1}_{+1}(\m_b);
\\
\label{MMSRSprimedet2}
\Rightarrow 
\;\;\;\;   \m_{c}
&=&1\,227^{-1}_{+18}(\nu)^{+11}_{-22}(\nu_f)^{-12}_{+12}(\als)^{+13}_{-13}(N_m)^{-1}_{+1}(\m_b)\blue{,}
\eea
where $\m_c$ is fixed from the RS masses at the scale $\nu=1.5$ GeV (again considering $\nu=2.5$ or $\nu=1$ GeV produces small changes in $\m_c$). Nicely enough the uncertainty associated to the bottom quark mass is quite small and so is the scheme dependence, i.e. the values obtained in \eqs{MMSRSdet2}{MMSRSprimedet2} are quite close. The uncertainties in $\m_c$ are dominated by the error of $\nu$.

Taking the average of the RS and RS' determinations of $\m_c$ we obtain
\be
\m_c  =  1223(33) \; {\rm MeV} \ .
\label{result2}
\ee 

For illustration, we show the value of the different orders in the perturbation series for the central value parameters:
\bea
 {\rm RS: \;}M_{B_c}-M_{\eta_b}/2&=& 
 ( 
1204 + 162 + 99 + 75 + 35
 ) \ {\rm MeV} \,, 
\\
{\rm RS': \;}M_{B_c}-M_{\eta_b}/2&=&
 ( 
1501 + 4 + 29 + 33 + 8
 ) \ {\rm MeV} \,, 
\\
m_{c, \RS}(1\;{\rm GeV})&=&
 ( 
1220 - 35 + 18 + 11 - 10
 ) \ {\rm MeV} \ ,
\\
m_{c, \RS'}(1\;{\rm GeV})&=&
 ( 
1227 + 183 + 62 + 30 - 1
 ) \ {\rm MeV} \ .
\eea
We see that the perturbative expansions are reasonably convergent. 

If we compare \eq{result1} with \eq{result2} the central values are almost identical. The determination from the $B_c$ has a slightly smaller error, even though this is strongly dependent on the scale variation we choose to take for the observables. Moreover, on theoretical grounds, the transfer energy in the $B_c$ (and $\eta_b$) is bigger than for charmonium. Therefore, we take this determination as more solid and choose it as our final determination. The RS and RS' masses determined from this value (and the $\m_c$ associated error) are collected in \Tab{Tab:mc}.

\begin{table}[h]
\begin{center}
\begin{tabular}{|c|c|c|c|}
  \cline{1-4}
\rule[-1ex]{0pt}{0pt} \rule{0pt}{3ex} $\nu_f$ &0.7 GeV & 1 GeV & 2 GeV \\
  \hline \hline
\rule[-2.5ex]{0pt}{0pt}\rule{0pt}{4ex}  $m_{c,\RS}$ (MeV)& 1\,326(41) &1\,209(41) & \;\,950(41) \\
 \hline
  \rule[-2.5ex]{0pt}{0pt}\rule{0pt}{4ex}  $m_{c,\RS'}$ (MeV) & 1\,592(41) & 1\,496(41) & 1\,316(41)  \\
  \hline
\end{tabular}
\caption{RS and RS' charm quark masses for different values of $\nu_f$.}\label{Tab:mc}
\end{center}
\end{table}


\begin{figure}[!htb]
	\begin{center}      
	\includegraphics[width=0.48\textwidth]{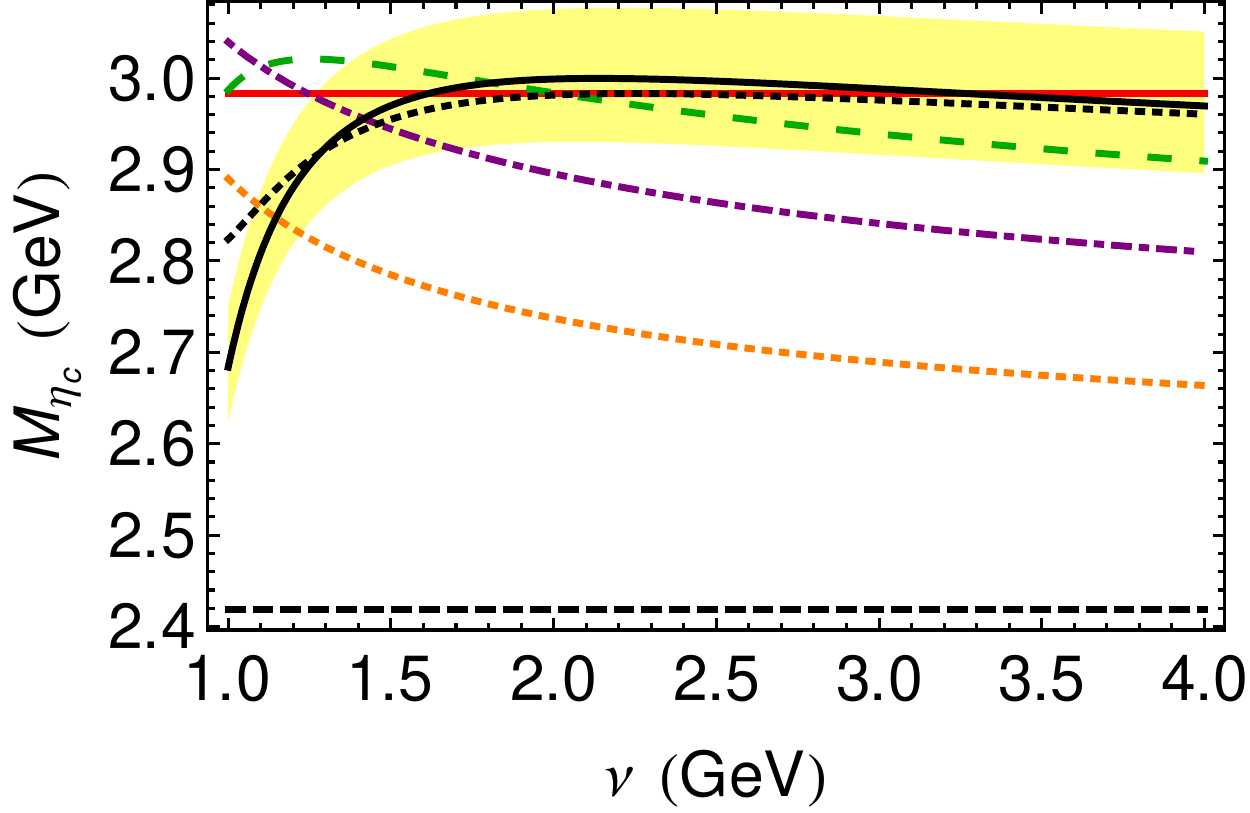}
	%
		\includegraphics[width=0.50\textwidth]{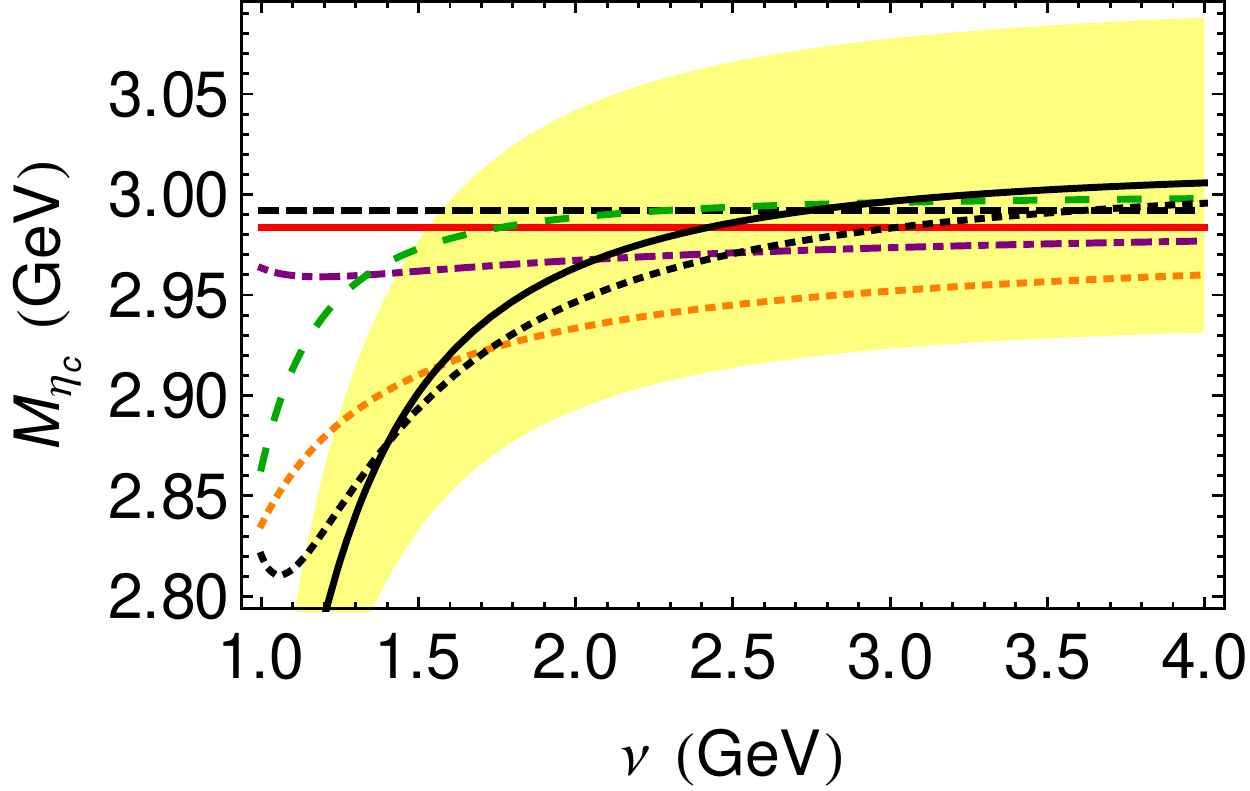}\\
		\includegraphics[width=0.48\textwidth]{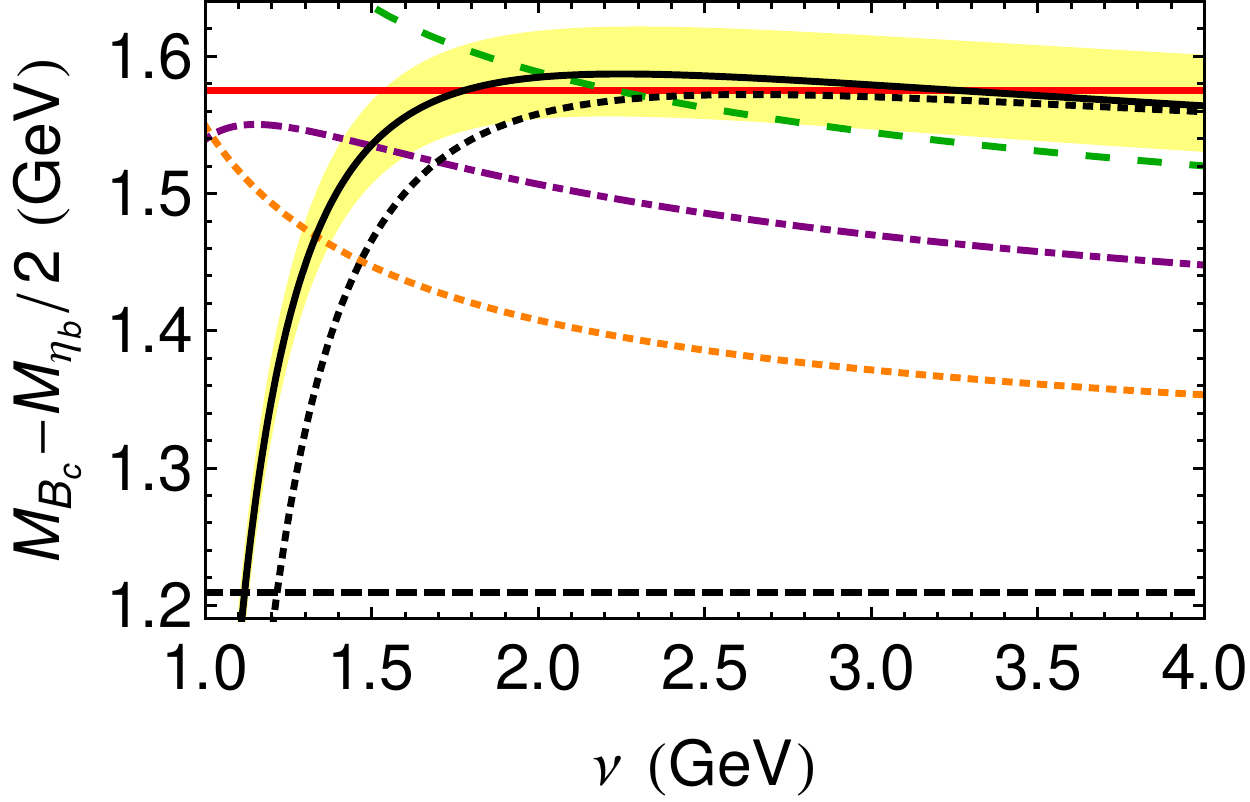}
	%
	\includegraphics[width=0.50\textwidth]{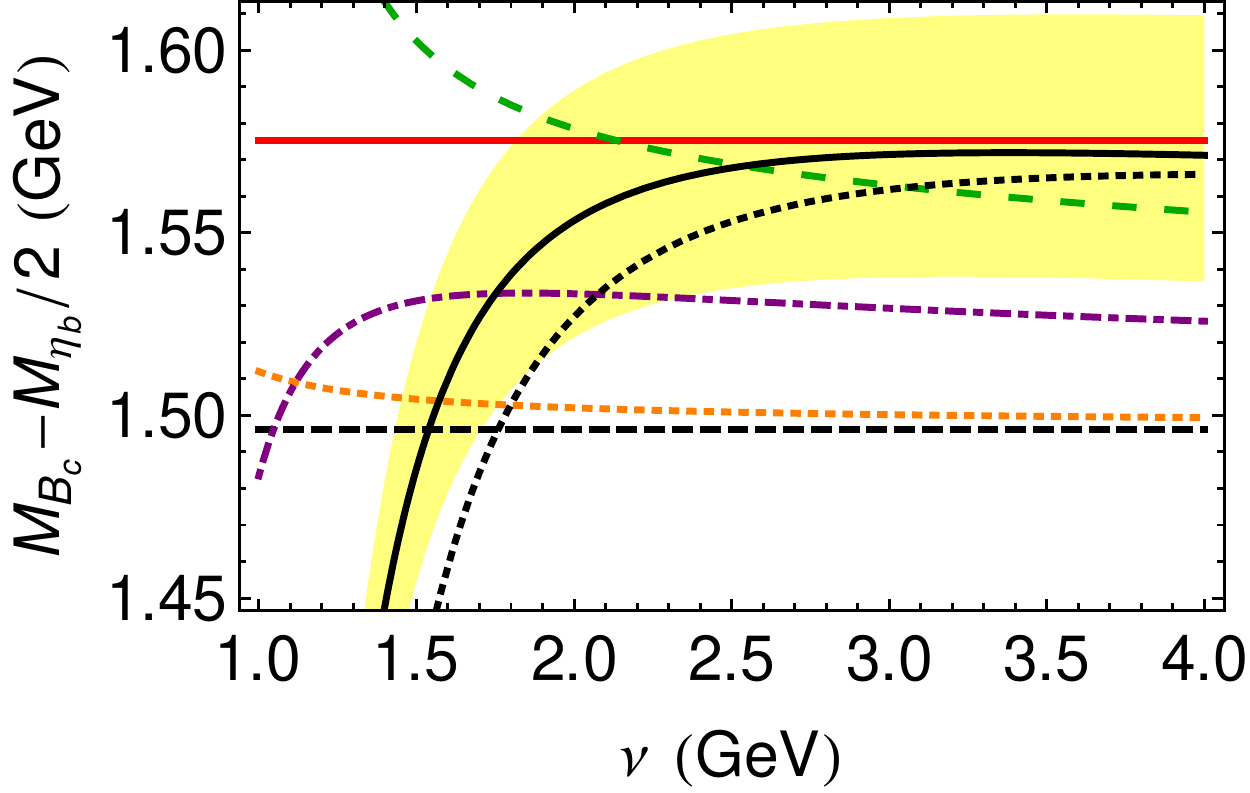}
\caption{Plots for $\nu_f=1$ GeV, $m_{c,\RS}=1.209^{-41}_{+41}(\m_c)$ GeV, $m_{c,\RS'}=1.496^{-41}_{+41}(\m_c)$ GeV.  Red line is experimental value, black dashed line is the sum of the masses. Orange dotted, purple dot-dashed and green long-dashed are the LO-NNLO contributions respectively. Solid black line is the N$^3$LO, and the dotted black line the N$^3$LO without $\delta E_{10}^{US}$. The yellow band corresponds to the error of
$m_{c,\RS/\RS'}$ associated to the uncertainty of $\m_c$.
 {\bf Upper left panel:} Plot of $M_{\eta_c}$ in RS scheme. 
{\bf Upper right panel:} Plot of $M_{\eta_c}$ in RS' scheme.
{\bf Lower left panel:} Plot of $M_{B_c}-M_{\eta_b}/2$ in RS scheme.
{\bf Lower right panel:} Plot of  $M_{B_c}-M_{\eta_b}/2$ in RS' scheme.
\label{Fig:etacBc}} 
\end{center}
\end{figure}

Let us now study in more detail the convergence of the perturbative series and the reliability of the error estimate. In  \fig{Fig:etacBc} we plot the theoretical prediction for $M_{\eta_c}$ and $M_{B_c}-M_{\eta_b}/2$ both in the RS and RS' as a function of $\nu$ truncating at different orders in the perturbative expansion. We emphasize that we use the same mass in the four cases: \eq{result2} (and also \eq{resultmbfinal} for $M_{B_c}-M_{\eta_b}/2$).  In the four cases the N$^3$LO predictions are in perfect accordance with experiment (for not too small scales). We also show the sensitivity of the results to variations of the charm mass in the yellow band of \fig{Fig:etacBc}. As expected, and needed for a precise determination of the heavy quark mass, we find such sensitivity to be very large.  In the plots in \fig{Fig:etacBc} we observe that the convergence is quite reasonable irrespectively of the scheme. 

Ultrasoft effects enter at  ${\cal O}(\als^5)$. As we did for the bottomonium, in order to quantify their importance  (or whether one can even compute their effects at weak coupling), we have also plotted the N$^3$LO prediction minus the pure ultrasoft contribution in \eq{EnlUS}. We observe that for the scales where our prediction is more robust (above 1.5-2 GeV) such contribution is small and well inside uncertainties. Actually, the fact that we obtain basically the same value for $\m_c$ from $M_{B_c}-M_{\eta_b}/2$ or $M_{\eta_c}$, hints that possible nonperturbative effects are strongly constrained. If ultrasoft effects cannot be computed in perturbation theory one can roughly estimate their size (for scaling purposes) to be of order $\lQ^3  \langle r^2 \rangle$. We would then expect to obtain different values for the charm mass from the two fits, since for the former the energy shift is
\be
\label{deltaENP1}
\delta E \sim \lQ^3\left( \langle r^2 \rangle_{Bc}-\frac{1}{2} \langle r^2 \rangle_{\eta_b} \right)
\ee
and for the latter 
\be
\label{deltaENP2}
\delta E \sim \lQ^3 \langle r^2 \rangle_{\eta_c}
\ .
\ee 
These two quantities are, in principle, different. Still, we basically find the same value for $\m_c$. This is compatible with assuming that either: 1) the ultrasoft scale can be dealt with in perturbation theory, 2) the nonperturbative effects are small, or 3) numerically, \eq{deltaENP1} is 1/2 of \eq{deltaENP2}. In the scenario 1), the leading nonperturbative effects can be written as an expansion in terms proportional to local condensates of higher and higher dimensionality. The dimension four corrections (proportional to the gluon condensate) were computed in \cite{Leutwyler:1980tn,Voloshin:1979uv}, the dimension six in \cite{Pineda:1996uk} and the dimension eight (for $l=0$) in \cite{Rauh:2018vsv}. We refrain from using those corrections in this paper until a clear signal of the associated asymptotic behavior of the perturbative expansion related with these condensates is seen. 

It is also comforting that the yellow error band associated to the uncertainty of $\m_c$ in \fig{Fig:etacBc} covers the difference between the NNLO and N$^3$LO results in the RS and in the RS' scheme for the whole scale variation in most cases. This difference is often used as an alternative determination of the error. Actually in the RS' scheme the difference between the different orders is quite small.   

As a final check, we have performed the fit to the $\Psi(1S)$ mass using the same setup. We find a shift of +41 MeV. This is a little bit less than one standard deviation of \eq{result1}, which we obtained from the $\eta_c$. This is consistent with the error estimate. This difference is related with the hyperfine splitting. It has been argued that such energy splitting is rather sensitive to the resummation of large  logarithms \cite{Kniehl:2003ap}, which we do not incorporate in this paper (we hope to come back to this issue in the future). On top of that, the hyperfine splitting of the $B_c$ minus $\Upsilon/2$ energy combination is expected to be much smaller than for charmonium. This reinforces in our choice of $M_{B_c}-M_{\eta_b}/2$ as the optimal observable to determine the charm mass.  

\begin{figure}[!htb]
	\begin{center}   
	\includegraphics[width=0.85\textwidth]{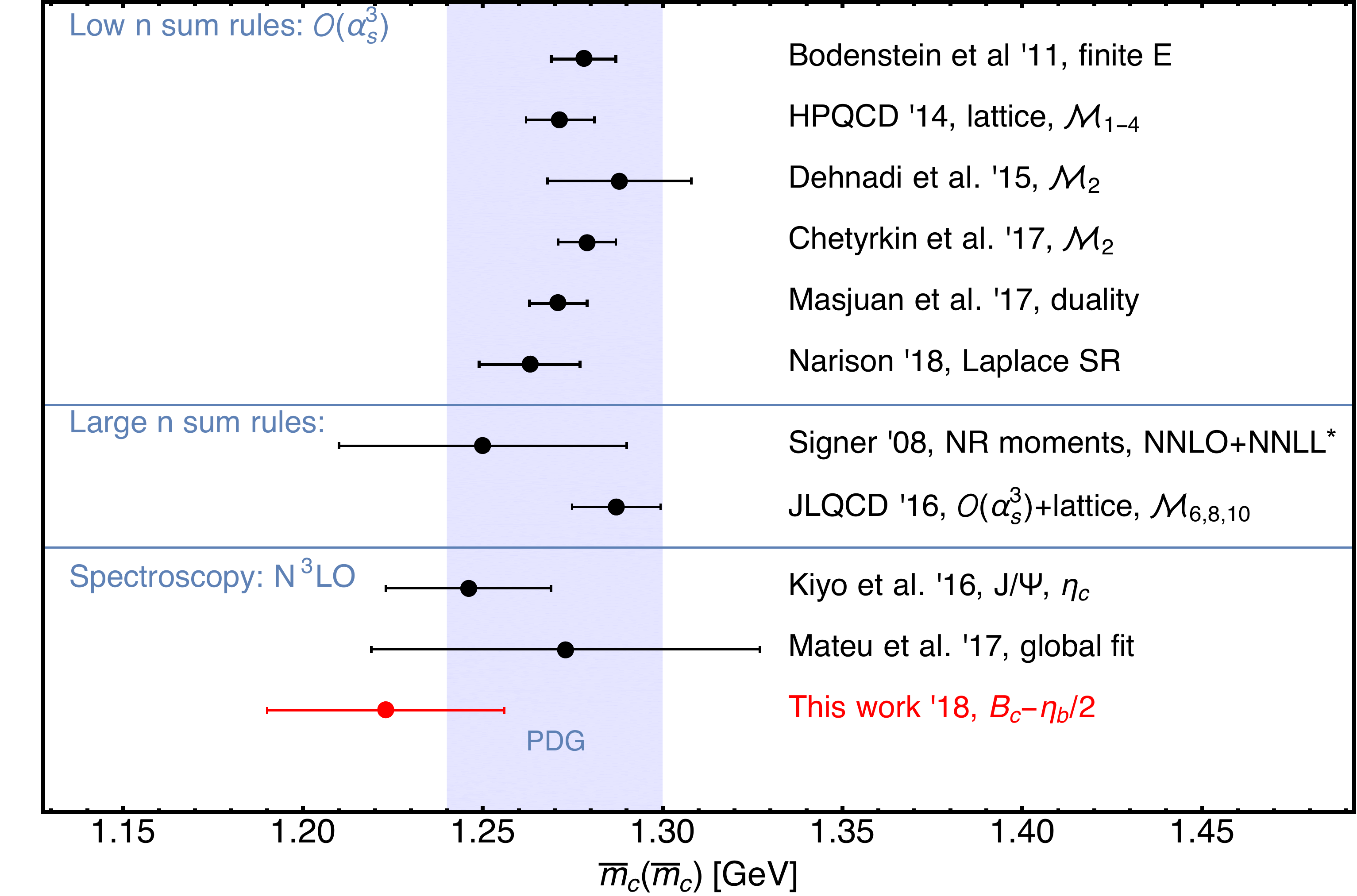}	      
\caption{$\m_{c}(\m_{c})$ determined from spectroscopy or from sum rules in the strict weak-coupling approximation. 
Chetyrkin et al. \cite{Chetyrkin:2017lif}, HPQCD14 \cite{Chakraborty:2014aca}, and Dehnadi et al. \cite{Dehnadi:2015fra} 
use N$^3$LO low n moments equated to the corresponding experimental moments evaluated using the available experimental information (masses and decays) supplemented with lattice information and/or assuming perturbation theory at high energies (in \cite{Erler:2016atg} this was indeed assumed since threshold using quark-hadron duality). 
Finite energy \cite{Bodenstein:2011ma} or Laplace \cite{Narison:2018dcr} sum rules have also been applied for low $n$. 
Signer \cite{Signer:2008da} uses NNLO and (a partial) NNLL expression for nonrelativistic (with Coulomb resummation) large $n$ moments.  JLQCD \cite{Nakayama:2016atf} uses N$^3$LO (without Coulomb resummation) for large $n$. 
Kiyo et al. \cite{Kiyo:2015ufa}, Mateu et al. \cite{Mateu:2017hlz} and this paper use N$^3$LO expressions for the heavy quarkonium masses. 
\label{Tabmc}}   
\end{center}
\end{figure}

We now compare our result with previous determinations of $\m_{c}(\m_{c})$ from spectroscopy or from sum rules. We show such comparison in Fig. \ref{Tabmc}. We find that all determinations are perfectly consistent with each other, but there is a small tension with determinations using low $n$ sum rules, where our number is in the low range. In any case, the difference is hardly statistically significant. Low $n$ sum rules determinations seems to give slightly larger values of the mass than large $n$ or spectroscopy determinations. We notice that the latter include the Coulomb resummation. Therefore, the physics they are sensitive to is different. Nevertheless, at this stage this difference is no statistically significant, and we are not in the position to 
 push this discussion further. It is also worth mentioning that, at present, there are no large $n$ N$^3$LO sum rules analyses (including the Coulomb resummation). 
 
We now turn our discussion to the comparison between different spectroscopy determinations. The discussion follows parallel to the discussion we had for the bottom mass. Our value is slightlty smaller than the determinations in \cite{Kiyo:2015ufa,Mateu:2017hlz} (but well within each other's uncertanties). Compared with the determinations of Mateu et al. \cite{Mateu:2017hlz}, a part of the difference can be traced back to the fact that in \cite{Mateu:2017hlz} the $J/\Psi$ plays a dominant role in the fit (which leads to a larger value for the charm mass than fits to the $\eta_c$ mass). The reason is that, even if both the vector and the pseudoscalar were considered in the fit, their method weights the most the particle that produces the smaller errror, which in their paper is the vector. It is worth mentioning that if one looks to the individual fits in \cite{Mateu:2017hlz} to the $\eta_c$ mass one finds a smaller value of $\m_{c}(\m_{c})$, quite close to ours. 
As at present all these determinations are consistent with each other, there is not a clear reason to prefer one versus the other. As we have already discussed before, the resummation of large logarithms may help to solve this problem. We hope to come back to this issue in the future. The other fact that produces a different central value is the choice of the factorization scale $\mu$ used to fix the central value of $\m_{c}(\m_{c})$. In this paper we have chosen a larger value of the factorization scale than in \cite{Mateu:2017hlz}. We somewhat feel that a smaller value of the factorization scale leads to a region where the perturbative series does not behave well enough. In this respect the factorization scales we choose are closer to the factorization scales used in \cite{Kiyo:2015ufa}. Actually, the difference with respect to the value obtained in \cite{Kiyo:2015ufa} can be traced back to the fact that in that reference an average of the determination of the charm mass from the $J/\Psi$ and the $\eta_c$ is made. Again pure determinations from the $\eta_c$ mass are closer to our numbers. It is also worth mentioning that the error of the $\m_{c}(\m_{c})$ in \cite{Kiyo:2015ufa} from the $J/\Psi$ or $\eta_c$ mass is quite similar (this is the reason why the value obtained in \cite{Kiyo:2015ufa} is more or less the aritmetic average between the values of $\m_{c}(\m_{c})$ obtained from the $J/\Psi$ and the $\eta_c$).

\section{Renormalon-free combinations and determination of $\als$}
\label{sec:RF}
\begin{figure}[!htb]
	\begin{center}
	\includegraphics[width=0.43\textwidth]{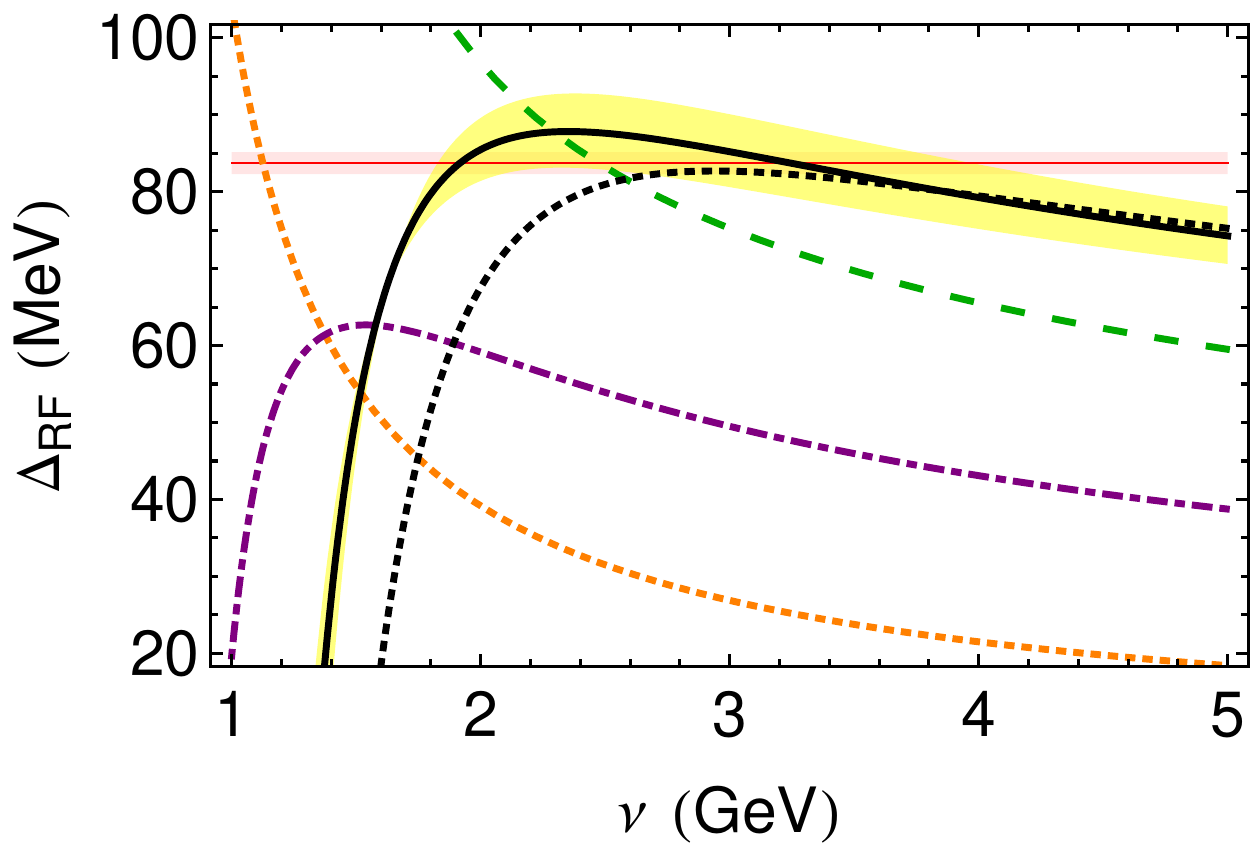}	      
	\includegraphics[width=0.43\textwidth]{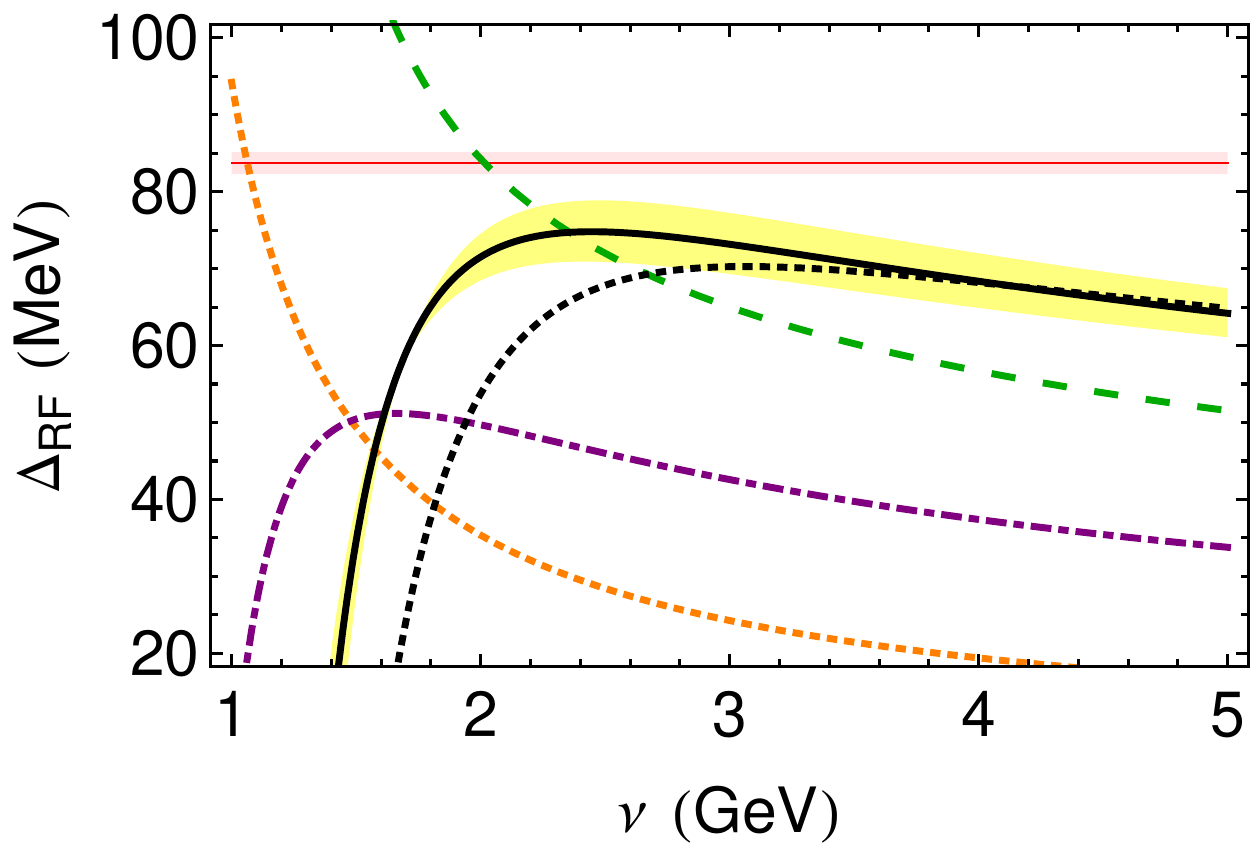}\\
	%
\caption{Plots of $\Delta_{\rm RF}=M_{B_c}-M_{\eta_b}/2-M_{\eta_c}/2$ with $\nu_f=1$ GeV, $m_{b,\RS}=4.618$ GeV, $m_{c,\RS}=1.209$ GeV, $m_{b,\RS'}=4.885$ GeV, $m_{c,\RS'}=1.496$ GeV. Red line is experimental value. Orange dotted, purple dot-dashed and green long-dashed are the LO-NNLO contributions respectively. Solid black line is the N$^3$LO, and the dotted black line the N$^3$LO without $\delta E_{10}^{US}$. The yellow band represents the error coming from $\als(M_z)=0.1184(12)$.
{\bf Left panel:} Plot of $\Delta_{\rm RF}$ in RS scheme.
{\bf Right panel:} Plot of $\Delta_{\rm RF}$ in RS' scheme. 
\label{Fig:etabetacBc}
}
\end{center}
\end{figure}
It is interesting to consider renormalon-free combinations. Particularly compelling is the combination 
\be
\label{MBcetabetac}
\Delta_{\rm RF}\equiv M_{B_c}-M_{\eta_b}/2-M_{\eta_c}/2\Big |^{\rm exp}= 83.7\pm 1.4\,{\rm MeV}
\ .
\ee
On the one hand we know its experimental value. On the other hand, for this observable, the leading renormalon ambiguity has cancelled\footnote{There is a residual renormalon dependence, $\delta V \sim \frac{{\bf p}^2}{m^2} \delta m_{\RS}$, which nevertheless cancels with the pole mass renormalon that appears in $V_{p^2}$, one of the relativistic potentials. Its numerical impact is very small.}. A possible $u=1$ renormalon ambiguity associated to the kinetic term also exactly cancels in this combination. Therefore, this observable is (potentially) particularly good for the analysis of ultrasoft effects, as any noise in the signal associated to these renormalons disappears. We emphasize that the  determinations of the bottom and charm mass in the previous section assume that perturbative computations are reliable at the ultrasoft scale, or at least that possible nonperturbative effects are small. Roughly, one can assume the nonperturbative effects of $\Delta_{\rm RF}$ to scale as (note though that this estimate looses the ultrasoft logarithms)
\be
\delta E \sim \lQ^3\left( \langle r^2 \rangle_{B_c}-\frac{1}{2} \langle r^2 \rangle_{\eta_b}-\frac{1}{2}\langle r^2 \rangle_{\eta_c} \right)
\,.
\ee
This combination of expectation values is not zero (though some cancellation is expected). 

We can compare \eq{MBcetabetac} with the theoretical prediction using the masses obtained in \Sec{Sec:mbot}. We show this in \fig{Fig:etabetacBc} for the RS and RS' schemes. We observe that the theory prediction is  consistent with experiment within the expected uncertainties. We take this plot as an indication that we can use perturbation theory for the ground state of the three systems. Looking at \fig{Fig:etabetacBc}, the incorporation of the perturbative ultrasoft contribution indeed seems to improve the agreement with data, even though the effect is small and can be perfectly masked by the uncertainties. 

\medskip

\begin{figure}[!htb]
	\begin{center}
		\includegraphics[width=0.43\textwidth]{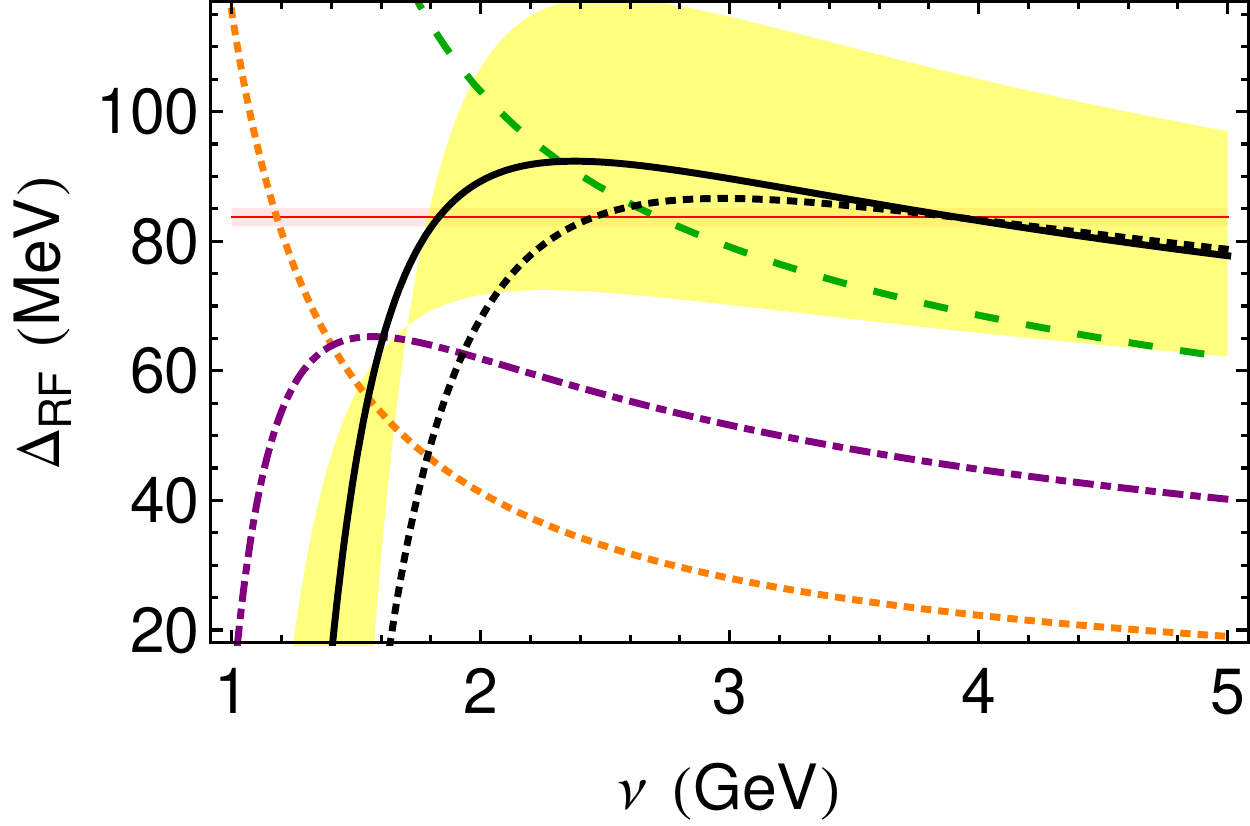}	      
	\includegraphics[width=0.43\textwidth]{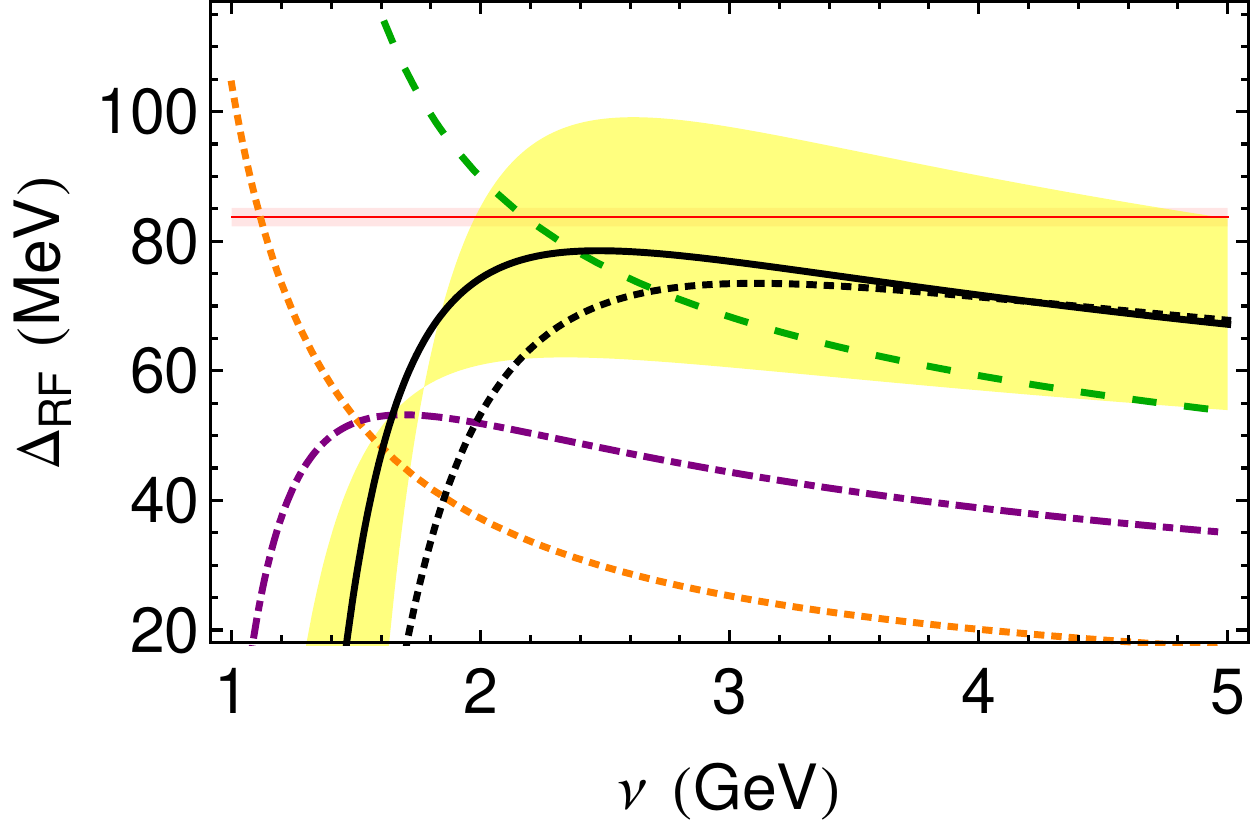}
	%
\caption{As in \fig{Fig:etabetacBc} changing the central value of $\als$ and the yellow band to the values in \eq{alscomb}. \label{Fig:RFalphacomb}}
\end{center}
\end{figure}

Indeed, assuming the weak-coupling approximation, $\Delta_{\rm RF}$ eliminates most of the mass dependence (both on the bottom and the charm quark masses). This makes this observable specially sensitive to $\als$. If we use it to fit $\als$ we obtain:
\bea
 {\rm RS: \;}\als(3 \ {\rm GeV})&=& 
[244.5^{+15}_{-1} (\nu)^{-1}_{+2} (\m_b)^{+4}_{-4} (\m_c)]10^{-3}=0.244(16)\label{alsRS}
 ,
 \\
\Rightarrow 
\;\;\;\;   \als(M_z)&=&0.1180^{+34}_{-36};
\\ 
 {\rm RS': \;}\als(3 \ {\rm GeV})&=& 
[260.1^{+17}_{+12} (\nu)^{-1}_{+1} (\m_b)^{+4}_{-4} (\m_c)]10^{-3}=0.260(17)\label{alsRSp}
,
 \\
\Rightarrow 
\;\;\;\;   \als(M_z)&=&0.1214^{+35}_{-38};
\eea
where we take $\nu=3^{+2}_{-1}$ GeV, $\nu_f=1$ GeV, and the masses from Tables~\ref{Tab:mb} and~\ref{Tab:mc}. In the last equality of \eqs{alsRS}{alsRSp} we have symmetrized the error. The lower value in the variation in $\nu$ is motivated by the fact that, at smaller scales, it is not possible to find a value of $\als$ that agrees with experiment. This hints that at smaller values of the scale, the observable does not behave as a (logarithmically modulated) polynomial in $\als$, with a small $\als$. The difference between the RS and RS' definitions is of the order of the error we obtain (unlike what happened for the mass determinations, where the difference was very small). The difference between the NNLO and N$^3$LO evaluation also gives an estimate for the error associated to higher order corrections. Again in this case it is of the order of the error. Note also that for the mass determinations our final error was bigger (and much bigger in some cases) than the difference between the N$^3$LO and the NNLO evaluation. Therefore, for the final error we take the minimum-maximum error of each determination in \eqs{alsRS}{alsRSp}:
\bea
 \als(M_z)&=&0.1195(53)
 \label{alscomb}
 \,.
\eea 
This is perfectly consistent with the world average albeit with larger errors. We show the plot with this value of $\als$ in \fig{Fig:RFalphacomb}, where the yellow band is the error quoted in \eq{alscomb}. We see that it is quite conservative. We leave possible improvements for the future. 

\medskip

\begin{figure}[!htb]
	\begin{center}      
	\includegraphics[width=0.43\textwidth]{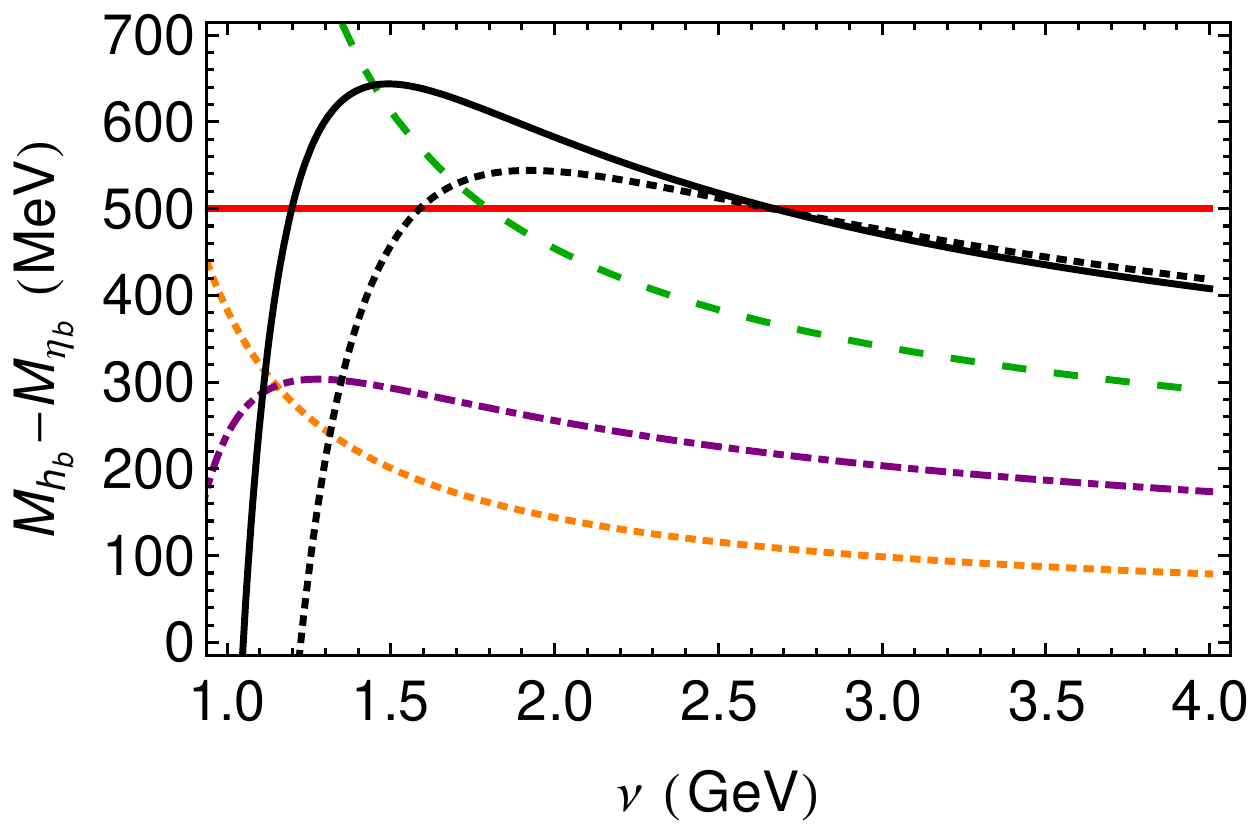}
	\includegraphics[width=0.43\textwidth]{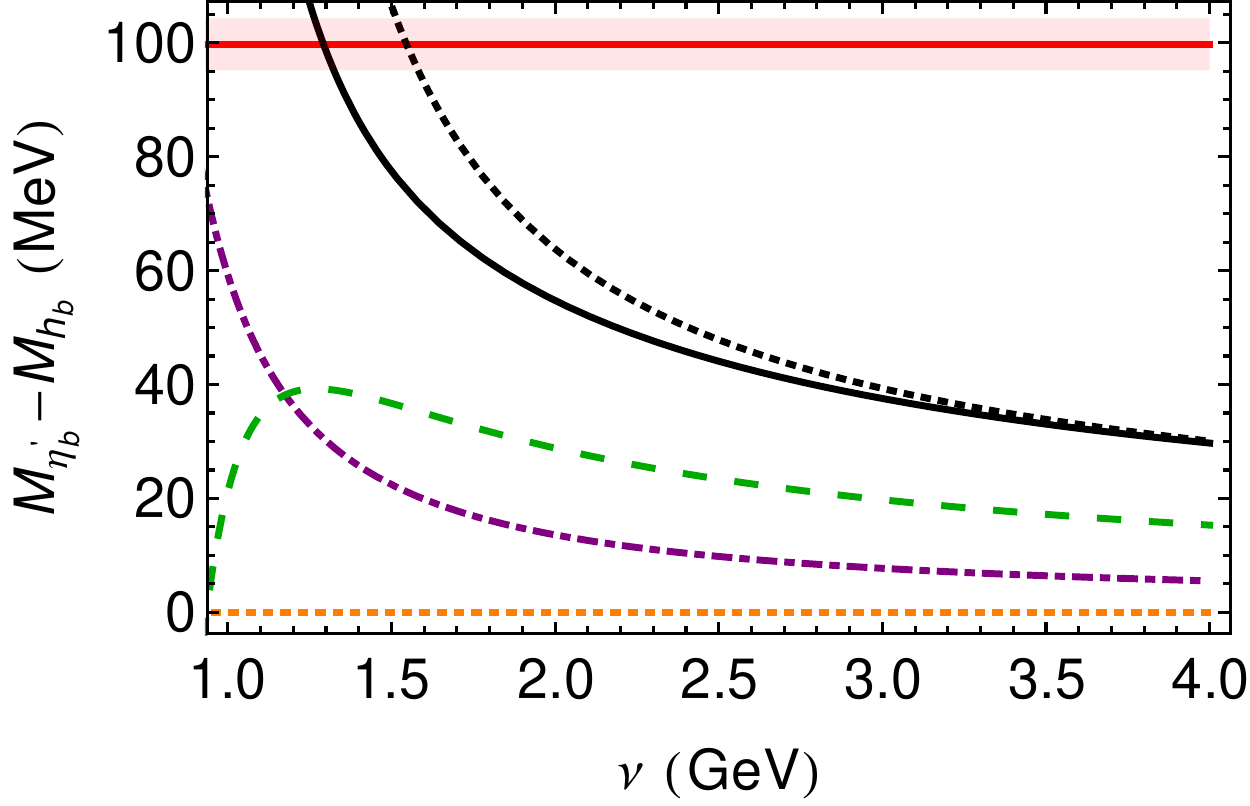}
	%
\caption{Plots for $n=2$ bottomonium states. The red line is experimental value. Orange dotted, purple dot-dashed and green long-dashed are the LO-NNLO contributions respectively. Solid black line is the N$^3$LO, and the dotted black line the N$^3$LO without $\delta E_{nl}^{US}$. {\bf Left panel:} Plot of $M_{h_b}-M_{\eta_b}$.
{\bf Right panel:} Plot of $M_{\eta'_b}-M_{h_b}$.
\label{Fig:SPwavebb}}   
\end{center}
\end{figure}

We now move from the ground state to the more shaky domain of excitations. Our aim is to see how reliable possible predictions of excited states are. We consider only $n=2$ bottomonium states, and energy differences among them and with the ground state. We show the plots in \fig{Fig:SPwavebb}, where we work in the RS' scheme with $\nu_f=1$ GeV. We see that even though they go in the right direction, the errors are large. The convergence is at most marginal. For the energy difference between the 1P and 1S state: $M_{h_b}-M_{\eta_b}$, 
the agreement with experiment is good for the N$^3$LO prediction, albeit with large uncertainties, only looking at the scale variation one finds $\sim \pm 150$ MeV. The $M_{\eta'_b}-M_{h_b}$ can be reproduced by the N$^3$LO result at $\nu \sim 1.5$ GeV but the convergence is not good and the renormalization scale dependence is rather large. In both cases, the uncertainties of the perturbative expansion are large in comparison with the ultrasoft contribution. 

\section{Higher orders}

In the previous sections we have seen that a strict (fixed-order) weak-coupling computation works well for the ground state of bottomonium, charmonium and the $B_c$. We have also explored the $n=2$ excitation for bottomonium. The convergence in this case was, at most, marginal. We now study a computational scheme that reorganizes the perturbative expansion such that it performs a selective sum of higher order corrections. 
We want to test if such scheme could improve/accelerate the convergence. In this method we incorporate the static potential exactly (to a given order) in the leading order Hamiltonian (the explicit $\nu$ dependence of the static potential appears at N$^3$LO order and partially cancels with the explicit $\nu$ dependence of \eq{EnlUS}, the ultrasoft correction): 
\be
\label{eq:Schroedinger}
\left[\frac{{\bf p}^2}{2m_r}+V_N^{(0)}(r;\nu)
\right]\phi^{(0)}_{nl}({\bf r})=E_{nl}^{(0)}\phi^{(0)}_{nl}({\bf r})
\,,
\ee
where the static potential will be approximated by a polynomial of 
order $N+1$, 
\begin{eqnarray}
V_N^{(0)}(r;\nu)
&=&
 -\frac{C_f\,\als(\nu)}{r}\,
\bigg\{1+\sum_{n=1}^{N}
\left(\frac{\als(\nu)}{4\pi}\right)^n a_n(\nu;r)\bigg\}
\,.
\label{VSDfo}
\end{eqnarray}
In principle we would like to take $N$ as large as possible (though we also want to explore the dependence on $N$). 
In practice we take the static potential at most up to N=3, i.e. up to
${\cal O}(\als^4)$ including also the leading ultrasoft corrections.
This is the order to which the coefficients $a_n$ are completely known:
\begin{eqnarray}
a_1(\nu;r)
&=&
a_1+2\beta_0\,\ln\left(\nu e^{\gamma_E} r\right)
\,,
\nonumber\\
a_2(\nu;r)
&=&
a_2 + \frac{\pi^2}{3}\beta_0^{\,2}
+\left(\,4a_1\beta_0+2\beta_1 \right)\,\ln\left(\nu e^{\gamma_E} r\right)\,
+4\beta_0^{\,2}\,\ln^2\left(\nu e^{\gamma_E} r\right)\,
\,,
\nonumber \\
a_3(\nu;r)
&=&
a_3+ a_1\beta_0^{\,2} \pi^2+
\frac{5\pi^2}{6}\beta_0\beta_1 +16\zeta_3\beta_0^{\,3}
\nonumber \\
&+&\bigg(2\pi^2\beta_0^{\,3} 
+ 6a_2\beta_0+4a_1\beta_1+2\beta_2
+\frac{16}{3}C_A^{\,3}\pi^2\bigg)\,
  \ln\left(\nu e^{\gamma_E} r\right)\,
\nonumber \\
&+&\bigg(12a_1\beta_0^{\,2}+10\beta_0\beta_1\bigg)\,
  \ln^2\left(\nu e^{\gamma_E} r\right)\,
+8\beta_0^{\,3}  \ln^3\left(\nu e^{\gamma_E} r\right)\,.
\label{eq:Vr}
\end{eqnarray}
The ${\cal O}(\als)$ term was computed in \rcite{Fischler:1977yf}, the ${\cal O}(\als^2)$  in \rcites{Schroder:1998vy,Peter:1996ig}, 
the ${\cal O}(\als^3)$ logarithmic term in \rcite{Brambilla:1999qa}, the light-flavour finite piece in 
\rcite{Smirnov:2008pn}, and the pure gluonic finite piece in \rcites{Anzai:2009tm,Smirnov:2009fh}.

We work in the following in the RS' scheme. Expressions for $\delta m_{\RS'}$ can be found in \rcite{Pineda:2013lta}. The static potential we will use then reads
\begin{equation}
\label{VRS}
V^{(0)}_{N,\RS'}(r;\nu)=
\,\left\{
\begin{array}{ll}
&
\displaystyle{
(V^{(0)}_{N}+2\delta m^{(N)}_{\RS'})|_{\nu=\nu}\equiv
 \sum_{n=0}^{N}V_{\RS',n}\als^{n+1}(\nu)
\qquad {\rm if} \quad r>\nu_r^{-1} }
\\
&
\displaystyle{
(V^{(0)}_{N}+2\delta m^{(N)}_{\RS'})|_{\nu=1/r}\equiv
 \sum_{n=0}^{N}V_{\RS',n}\als^{n+1}(1/r)
\qquad {\rm if} \quad r<\nu_r^{-1}. }
\end{array} \right.
\end{equation}

\eq{VRS} encodes all the possible relevant limits:

{\bf (a)}.
 The case $\nu_r=\infty$, $\nu_f=0$ is nothing but the on-shell static potential at fixed order, i.e. \eq{VSDfo}. 
 Note that the $N=0$ case reduces to a standard computation with a Coulomb potential, for which we can compare with analytic results for the matrix elements. We use this fact to check our numerical solutions of the Schr\"odinger equation. 
 
 {\bf (b)}.
The case $\nu_r=\infty$ (with finite non-zero $\nu_f$) is nothing but adding an $r$-independent constant to the static potential (see the discussion in \rcite{Pineda:2002se}). 

{\bf (c)}. The case $\nu_r=$finite (and, for consistency, $\nu_r \geq \nu_f$). We expect this case to improve over the previous results, as it incorporates the correct (logarithmically modulated) short distance behavior of the potential. This has to be done with care in order not to spoil the 
renormalon cancellation. For this purpose it is compulsory to keep a finite, non-vanishing, $\nu_f$,  otherwise the renormalon cancellation is not achieved order by order in $N$, as it was discussed in detail in 
\rcite{Pineda:2002se}. We have explored the effect of different values of $\nu_f$ in our analysis. Large values of $\nu_f$ imply a large infrared cutoff. In this way our scheme becomes closer to an $\MS$-like scheme. Such schemes still achieve renormalon cancellation, yet they jeopardize the power counting, as the residual mass $\delta m_{\RS'}$ does not count as $mv^2$. As a consequence, consecutive terms of the perturbative series become bigger. Therefore, we prefer values of $\nu_f$ as low as possible, with the constraints that one should still obtain the renormalon cancellation, and that it is still possible to perform the expansion in powers of $\als$. 

\begin{figure}[!htb]
	\begin{center}      
	\includegraphics[width=0.56\textwidth]{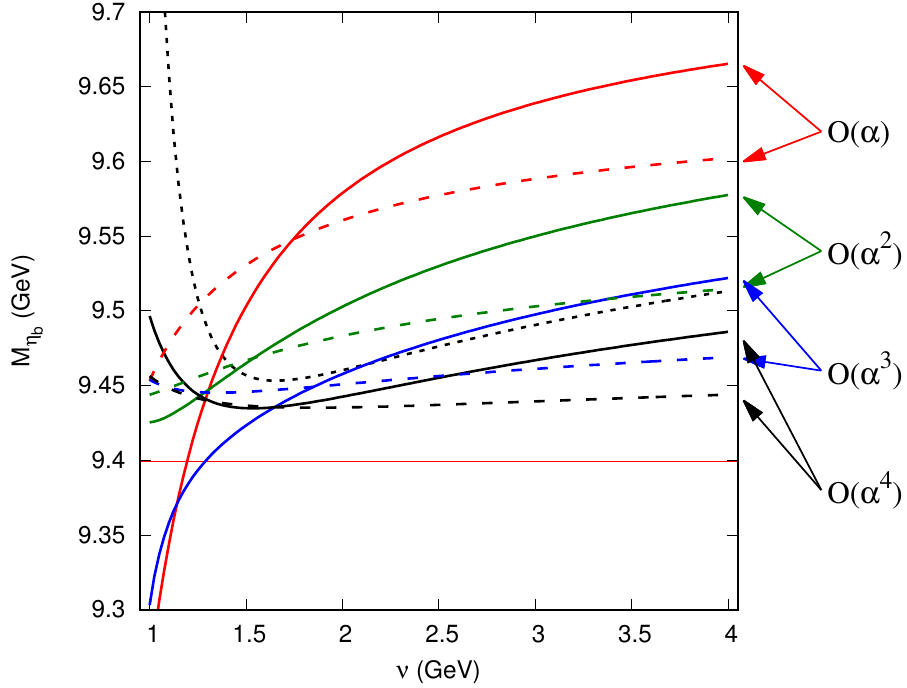}
	%
\caption{Plot of $2m_b+E_{10}^{(0)}$ for bottomonium using $V^{(0)}_{N,\RS'}$ for $N=0$, 1, 2, 3 with $\nu_f=1$ GeV. The red horizontal line is the experimental value. Continuous lines are evaluated with $\nu_r=\infty$ and dashed lines with $\nu_r=1$ GeV. The dotted line stands for the $N=3$ evaluation in strict weak coupling (truncating the sum at ${\cal O}(m\als^5)$). 
\label{Fig:Staticbb}}   
\end{center}
\end{figure}
 
\begin{figure}[!htb]
	\begin{center}
	\includegraphics[width=0.43\textwidth]{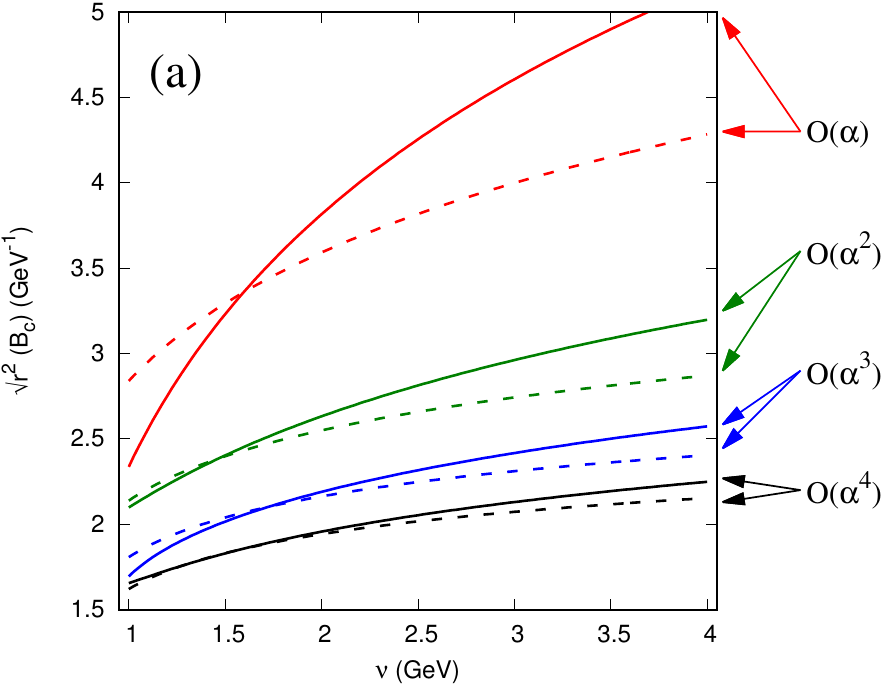}	      
	\includegraphics[width=0.43\textwidth]{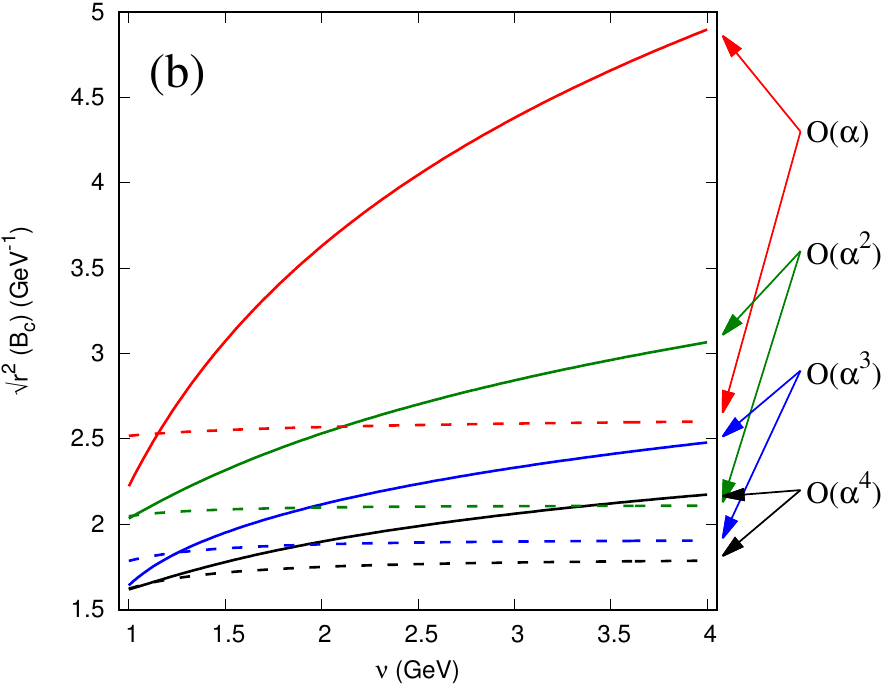}\\
	\includegraphics[width=0.43\textwidth]{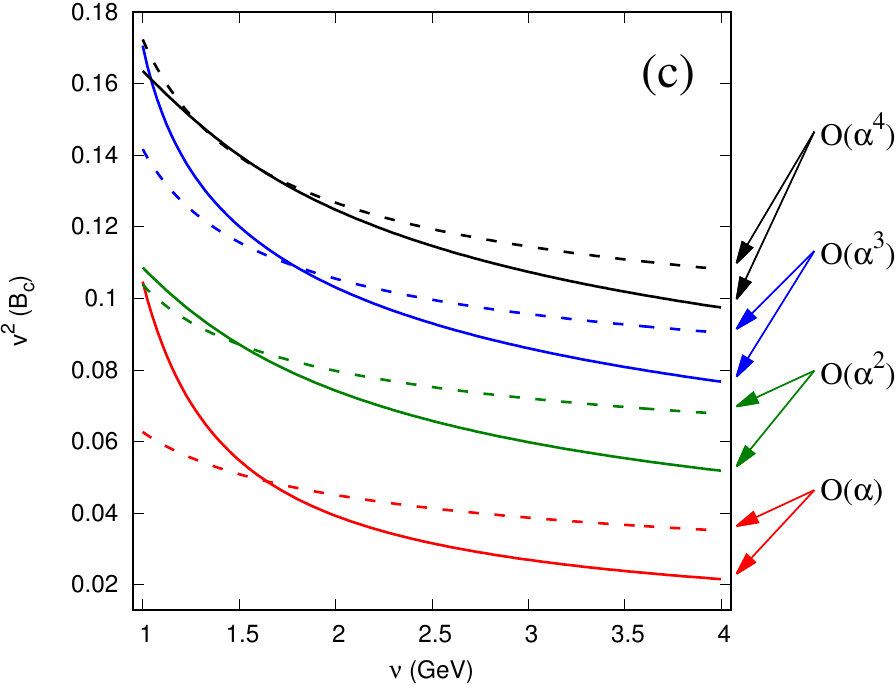}	      
	\includegraphics[width=0.43\textwidth]{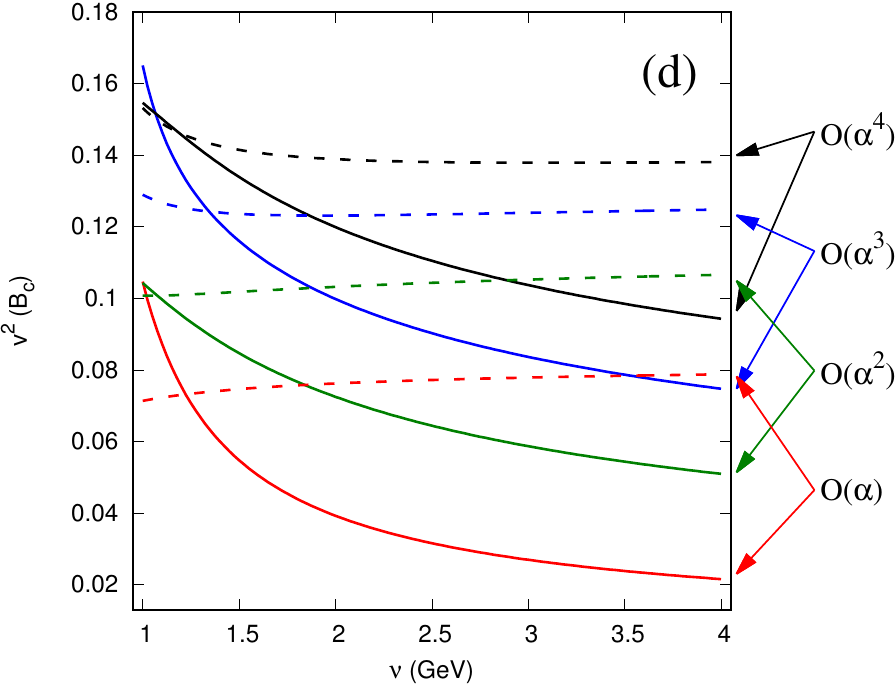}
	%
\caption{Plot of the radius and $v^2$ of the $B_c$ in the RS' scheme with $\nu_f=1$ GeV and $\nu_r=\infty$ (left panels), and with 
$\nu_f= \nu_r=0.7$ GeV (right panels). They are obtained from solving \eq{eq:Schroedinger} for $N=0$, 1, 2, 3. Continuous lines are evaluated with $\nu_r=\infty$ and dashed lines with $\nu_r=1$ GeV.
\label{Fig:radius}
}   
\end{center}
\end{figure}

$E_{nl}^{(0)}$ correctly incorporates the N$^N$LO corrections to the spectrum associated to the static potentials. It also includes higher order corrections (those generated by the iteration of the static potential). In order for this computational scheme to make sense, it first requires that the $N \rightarrow \infty$ converges, or at least that the error is small compared with the relativistic correction. This is indeed so for the bottomonium ground state. We show the result of the computation of $E_{10}^{(0)}$ for bottomonium in \fig{Fig:Staticbb}. We see that the convergence is good. There is an error associated to the computation, which we estimate as the difference between the last two terms in the expansion.  Setting $\nu_r=1$ GeV makes the result more stable under scale variations (it can even be made more stable if we take $\nu_f=\nu_r=0.7$ GeV). 
Typically, the $\nu_r=\infty$ and the $\nu_r=1$ GeV evaluation agree in the 1-2 GeV $\nu$ range.  
Moreover, the differences with the strict fixed-order computations are not large. This allows us to take\footnote{Strictly speaking the LO would be $2 m_b$.} $E_{nl}^{(0)}$ as the (more or less well defined) LO on top of which one can incorporate relativistic and/or ultrasoft corrections. A similar analysis for $B_c$, $\eta_c$ and bottomonium $n=2$ with $\nu_f=1$ GeV does not show such convergent behavior, though it does with $\nu_f=0.7$ GeV. If we instead consider renormalon-free observables, they typically show a better behavior. Therefore, these are the only combinations we consider in this paper and postpone a more detailed study of renormalon-dependent observables to future work. In this paper we usually work with $\nu_r=1\text{ or }\infty$ GeV but explore in some cases $\nu_r=0.7$ GeV. An earlier discussion, including also the charmonium ground state, can already be found in \rcite{Pineda:2013lta}. Nevertheless, the $B_c$ system has never been considered before. Therefore, following \rcite{Pineda:2013lta}, we profit to get an estimate of the electromagnetic radius and the mean velocity of the $B_c$. The analysis can be found in \fig{Fig:radius}. We find 
\be
\sqrt{\langle r^2 \rangle_{B_c}} \simeq 1.7 \; {\rm GeV}^{-1}\,, \qquad v \equiv \sqrt{\langle \frac{{\bf p}^2}{4m_r^2} \rangle_{B_c}} 
\simeq 0.37
\,.
\ee
The radius of the ground state $B_c$ is bigger than the radius of the ground state of bottomonium, and so is the mean velocity of the constituents. Overall, the convergence is good.

Once we have our new LO (the solution from the static potential), we can consider the incorporation of the relativistic and ultrasoft corrections. With the accuracy of this work, we only have to take the expectation value of $\delta V$ where 
\be
\delta V=V_s-V^{(0)}
\ee
stands for the relativistic potential ($V_s$ is the total singlet potential) that contributes up to N$^3$LO 
and also add the ultrasoft correction from \eq{EnlUS}. Overall the mass of the bound states reads
\be
\label{Mnlfull}
M(n,l,j)=m_1+m_2+E_{nl}^{(0)}+{}^{(0)}\langle n,l|\delta V|n,l \rangle^{(0)}+\delta E_{nl}^{US}
\,,
\ee
where $E_{nl}^{(0)}$ counts as $v^2$, ${}^{(0)}\langle n,l|\delta V|n,l \rangle^{(0)}$ counts as $v^4$ (including also $v^4 \als$ corrections) and $\delta E_{nl}^{US}$ as $v^5$. 
\eq{Mnlfull} is numerically correct with N$^3$LO precision and incorporates extra subleading terms (albeit in an incomplete way). 

This computational scheme resums a subset of subleading corrections in the hope that they would account for the bulk of such subleading terms. This could be achievable if the higher order corrections that we infer from our knowledge of the static potential are responsible of the leading corrections.
This computational scheme has already been applied in \rcites{Pineda:2013lta,Kiyo:2010jm}, where indeed an improved description of experiment was
 observed (see these references for extra details). 
 
The expressions we use for the relativistic potential (valid also in the unequal mass case) are taken from \rcite{Peset:2015vvi}, which uses results from \rcites{Pineda:1998kj,Kniehl:2001ju}. We can use any of the bases for the potentials presented in that paper, which were referred as: Wilson, onshell, Coulomb or Feynman. At strict N$^3$LO they all yield the same result. Since the computational scheme we implement in this section partially resums higher orders some dependence on the basis of potentials shows up. We have checked that, for the set of bases we consider, the dependence is quite small.

The computation of the relativistic corrections opens new issues compared with the static potential. In the case of the static potential the natural scale is $\nu \sim 1/r$, except in the ${\cal O}(\als^4)$ term where also the ultrasoft scale $\nu_{us}$ appears. The case of the relativistic potentials is quite different. They are much more dependent on the hard, and above all, the ultrasoft scale. Moreover, in order for the computation with the static potential to be a more or less reasonable approximation we need to have at least three or more terms (also important is the resummation of soft logarithms). For the case of the relativistic potentials, we have at most two terms. This, together with a much stronger scale dependence can make that inefficiencies of the description of the relativistic potentials get amplified when computing the expectation values. If, for instance, we consider $M_{\eta_b}$, we find that the relativistic corrections obtained with this computational scheme and with the strict fixed-order computation are very different. For the former the relativistic corrections are much larger than for the latter. This is different to the case of the static potential, where both approaches show comparatively small differences. This could be due to the fact that, for the relativistic corrections, we do not have enough terms (two at most) or that the unknown scale dependence of higher orders is very important and we are not describing it well enough. Actually, preliminary computations suggest that the resummation of logarithms is indeed important. We postpone this discussion for future research.   

\begin{figure}[!htb]
	\begin{center}
	\includegraphics[width=0.46\textwidth]{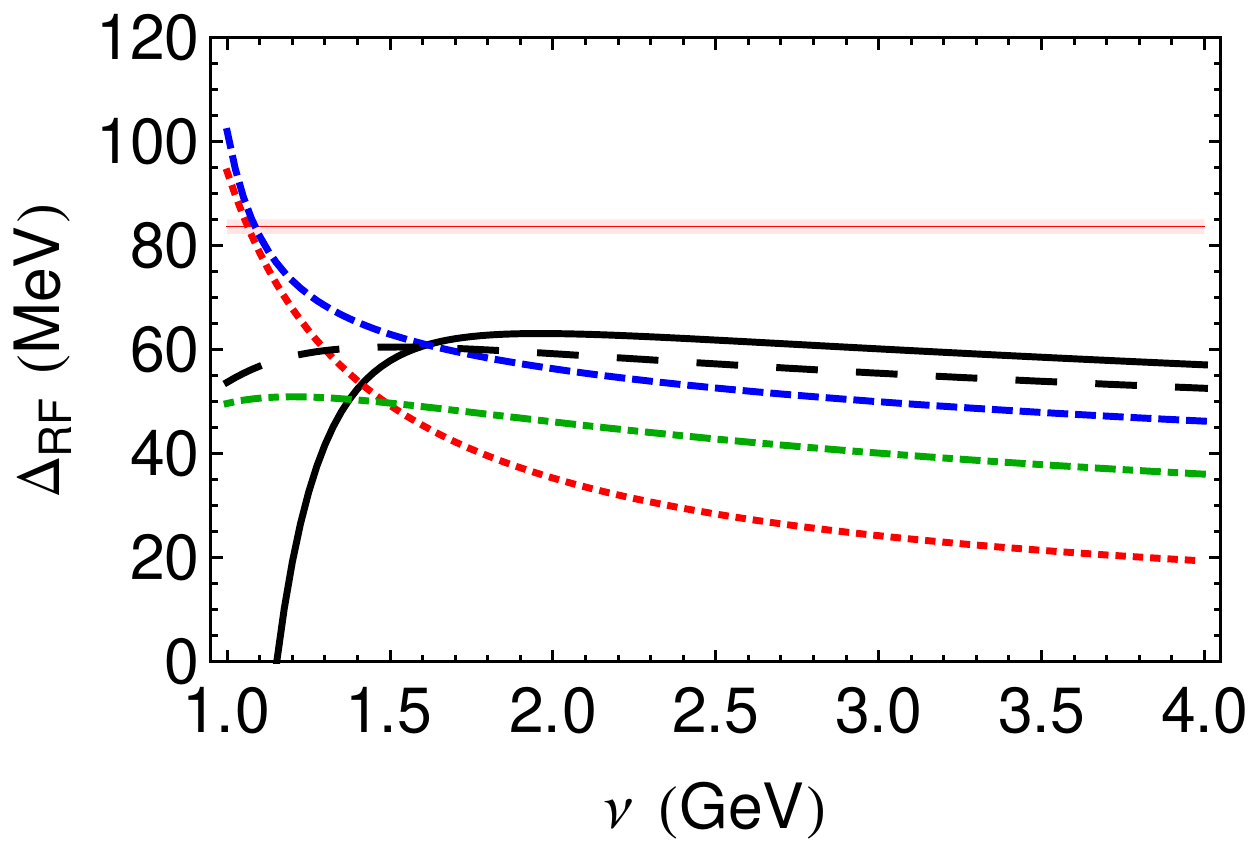}	      
	\includegraphics[width=0.46\textwidth]{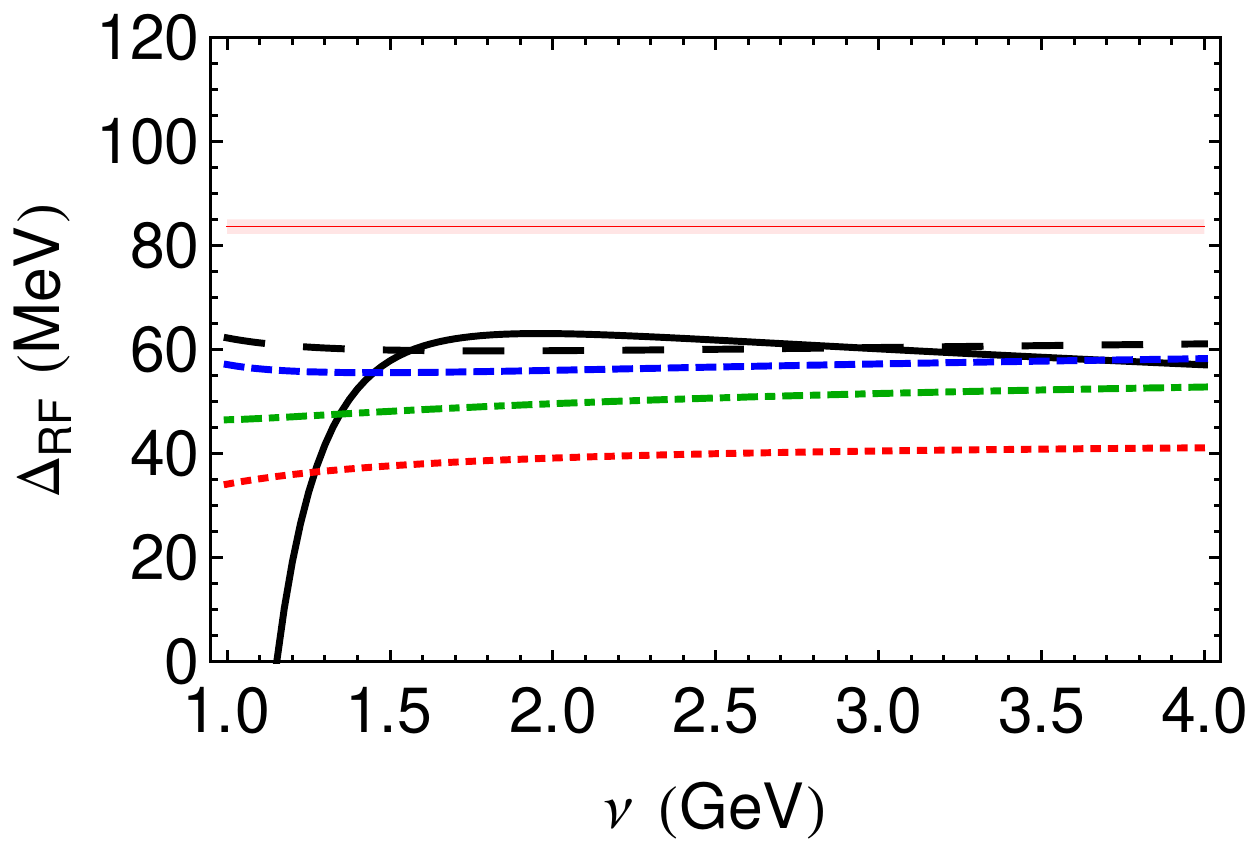}\\
	\includegraphics[width=0.46\textwidth]{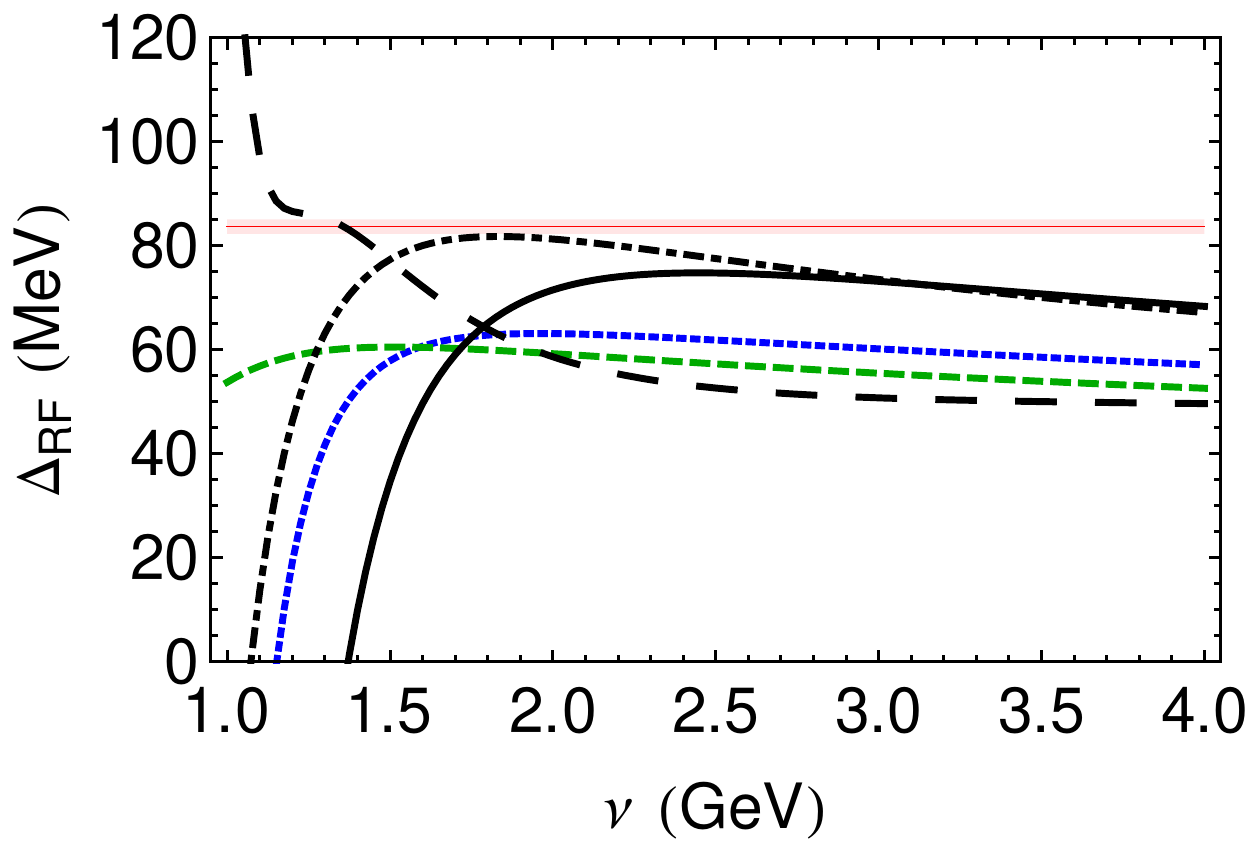}	      
	\includegraphics[width=0.46\textwidth]{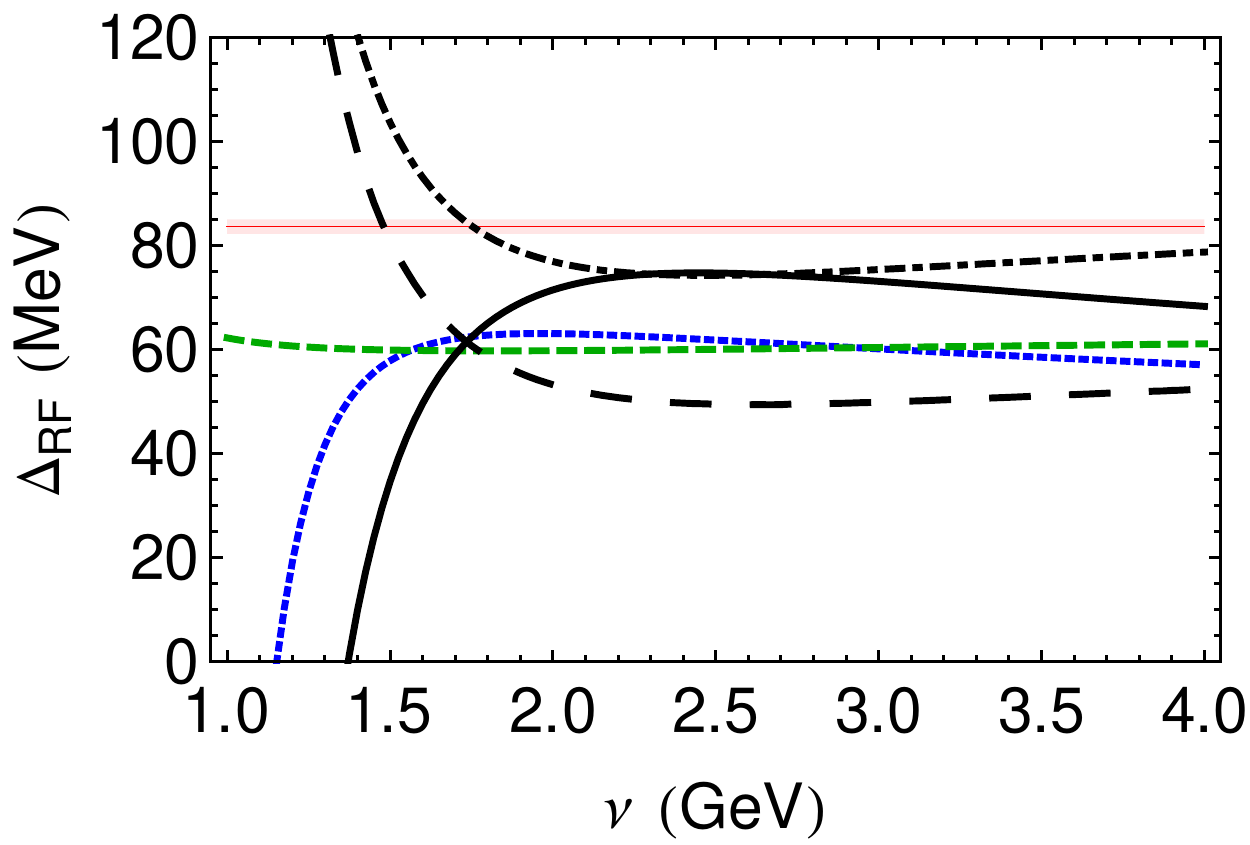}\\
	%
\caption{  
Plot of $\Delta_{\rm RF}=M_{B_c}-M_{\eta_b}/2-M_{\eta_c}/2$ in MeV in the RS' scheme for $\nu_f=1$ GeV, $m_{b,\RS'}=4.885$ GeV, $m_{c,\RS'}=1.496$ GeV. Red band is the experimental value.
{\bf Upper left panel:} $\nu_r=\infty$ GeV. Red dotted, green dot-dashed, blue dashed and black long-dashed are the results obtained solving \eq{eq:Schroedinger} with the static potential truncated at N=0,1,2,3, respectively. Solid black line is the $N=3$ previous result in strict fixed-order perturbation theory. 
{\bf Upper right panel:} As in the upper left panel but with $\nu_r=1$ GeV.   
{\bf Lower left panel:} $\nu_r=\infty$ GeV. Blue dotted line corresponds to the solid black line of the above figure, green short-dashed line corresponds to the long-dashed black line of the above figure. The other lines incorporate the relativistic and ultrasoft corrections. Solid black line is the full result in strict  fixed-order perturbation theory (solid black line in \fig{Fig:etabetacBc}).  The long-dashed black line corresponds to \eq{Mnlfull} evaluating the relativistic corrections with the wave function obtained from the $N=3$ static potential. The dot-dashed black line corresponds to \eq{Mnlfull} evaluating the ${\cal O}(mv^4)$ relativistic corrections with the wave function obtained from the $N=1$ static potential and the ${\cal O}(mv^4\als)$ relativistic corrections with the wave function obtained from the $N=0$ static potential.
{\bf Lower right panel:} As in the lower left panel but with $\nu_r=1$ GeV. 
\label{Fig:etabetacBcmur}
}   
\end{center}
\end{figure}
To gain further insight we study in detail $M_{B_c}-M_{\eta_b}/2-M_{\eta_c}/2$, which is a renormalon-free observable. We show our results in \fig{Fig:etabetacBcmur}, where we plot our predictions for $\nu_r=1$ GeV or $\infty$.  We observe that the LO solution is nicely convergent in $N$ in both cases, and also quite similar to the strict fixed-order result. This is quite remarkable as the individual contributions to the static potential do not converge that well for the case of the $\eta_b$, and certainly less for the case of the $\eta_c$ and the $B_c$ (nor they individually agree with the strict fixed-order computation, in particular in the last two cases). Overall, the static potential gives an energy shift $\sim 60$ MeV. 
For both values of $\nu_r$ the difference with the strict fixed-order computation and among themselves is quite small for scales bigger than 1.5 GeV, for scales below the result is quite sensitive to the partial incorporation of higher order corrections. Setting $\nu_r=1$ GeV makes the result quite scale independent. 

The relativistic corrections in \fig{Fig:etabetacBcmur} show a worse behavior. If we compute the expectation values of the relativistic potentials using the wave function obtained solving the Schrodinger equation with the N=3 static potential, the result is different from the result obtained from the strict fixed-order computation by around $\sim 20$ MeV, as we can see in the plot. One can basically recover the strict fixed-order computation by evaluating the relativistic correction with the matrix elements obtained solving \eq{eq:Schroedinger} with $N=1$ for the leading relativistic correction and with $N=0$ for the subleading relativistic corrections. Setting a smaller value of $\nu_r \sim 0.7$ GeV further deteriorates the agreement with experiment.   
We do not have a clear explanation for this fact. As we already mentioned before, we conjecture that it is fundamental to properly account for the different scales, as they may significantly affect the strength of the relativistic correction. Nevertheless, it is also worth to mention that we are talking about a difference with the strict fixed-order result of order $\sim 20$ MeV. This is much smaller than the differences we found for the relativistic corrections to the energies of each individual state. This strong cancellation may indicate that effects such as accounting for the different scales or truncating the series of the relativistic potentials produce smaller errors for the observable $\Delta_{\rm RF}$. Actually, $\sim 20$ MeV is of the order of the difference between the RS and RS' computation at strict N$^3$LO fixed-order. Finally, we explore the effect of the ultrasoft term $\delta E^{US}_{10}$. We find it is comparatively small for scales bigger than 1.5 GeV.

\begin{figure}[!htb]
	\begin{center}
	\includegraphics[width=0.46\textwidth]{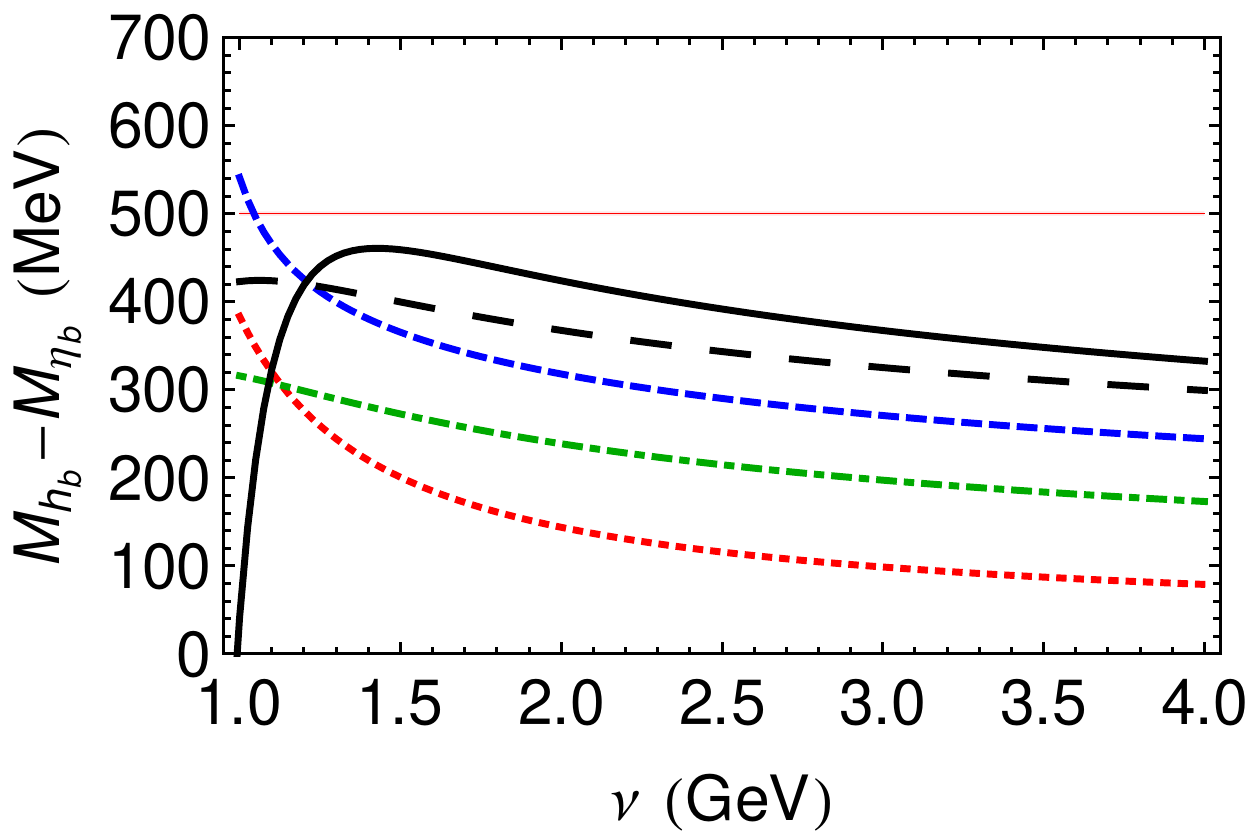}	      
	\includegraphics[width=0.46\textwidth]{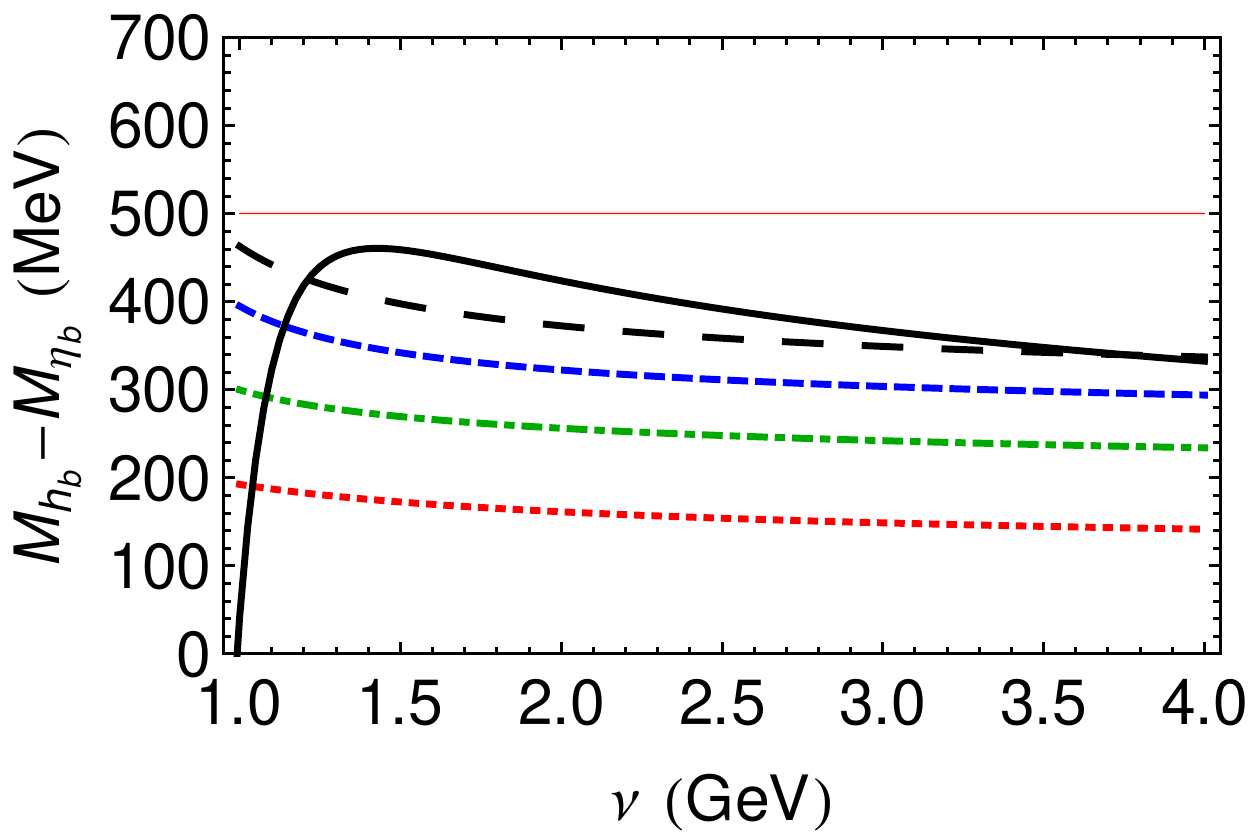}\\
	\includegraphics[width=0.46\textwidth]{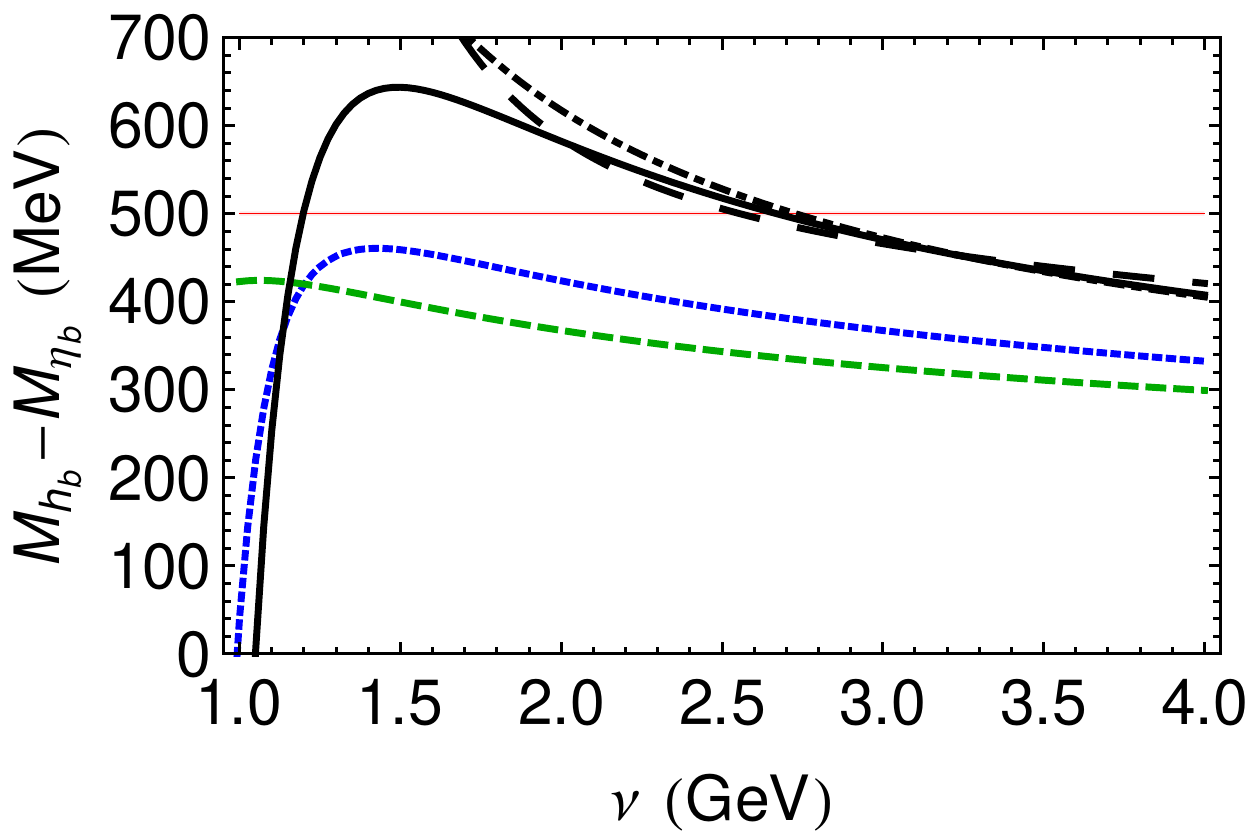}	      
	\includegraphics[width=0.46\textwidth]{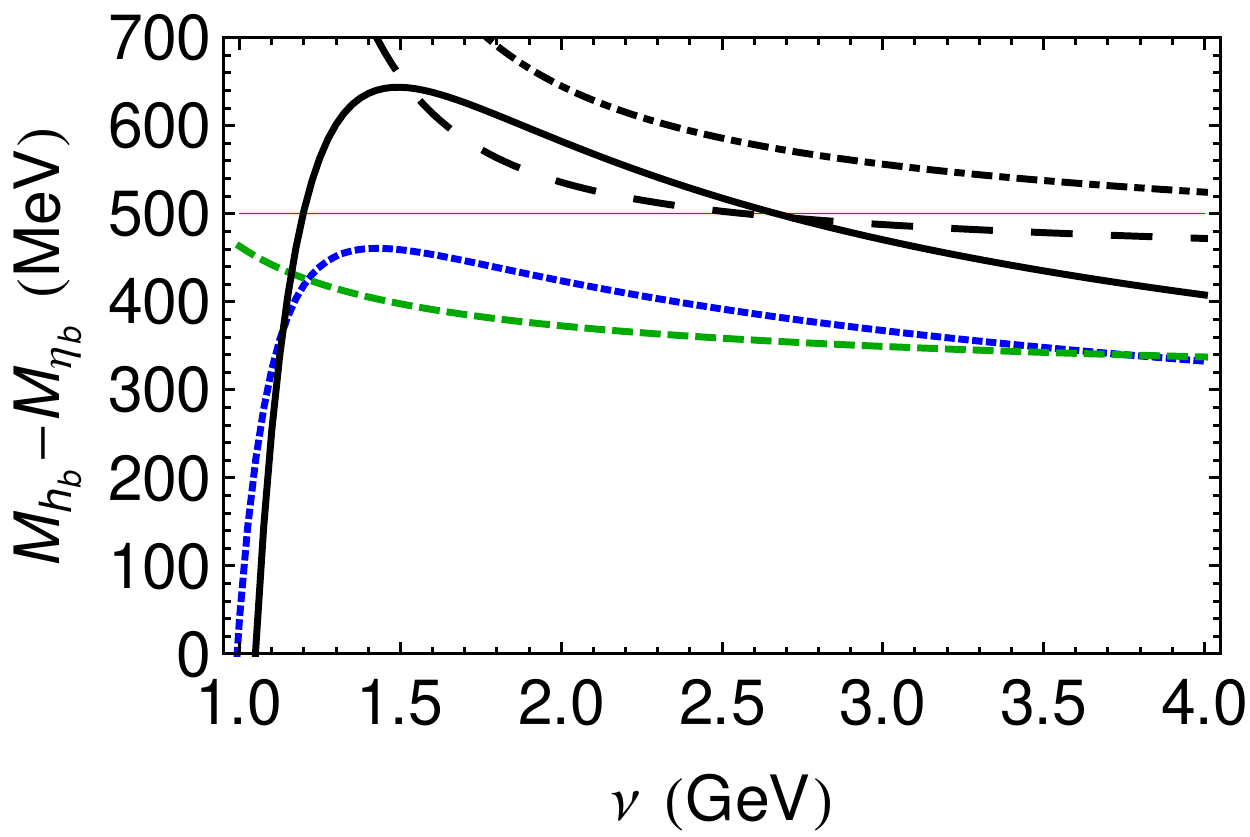}\\
	%
\caption{
Plot of $M_{h_b}-M_{\eta_b}$ in MeV in RS' scheme for $\nu_f=1$ GeV, $m_{b,\RS'}=4.885$ GeV. Red band is the experimental value.
{\bf Upper left panel:} $\nu_r=\infty$ GeV. Red dotted, green dot-dashed, blue dashed and black long-dashed are the results obtained solving \eq{eq:Schroedinger} with the static potential truncated at N=0,1,2,3, respectively. Solid black line is the $N=3$ previous result in strict fixed-order perturbation theory. 
{\bf Upper right panel:} As in the upper left panel but with $\nu_r=1$ GeV.   
{\bf Lower left panel:} $\nu_r=\infty$ GeV. Blue dotted line corresponds to the solid black line of the above figure, green short-dashed line corresponds to the long-dashed black line of the above figure. The other lines incorporate the relativistic and ultrasoft corrections. Solid black line is the full result in strict fixed-order perturbation theory (solid black line in the left panel in \fig{Fig:SPwavebb}).  The long-dashed black line corresponds to \eq{Mnlfull} evaluating the relativistic corrections with the wave function obtained solving the Schroedinger equation with the $N=3$ static potential. The dot-dashed black line corresponds to \eq{Mnlfull} evaluating the ${\cal O}(mv^4)$ relativistic corrections with the wave function obtained from the $N=1$ static potential and the ${\cal O}(mv^4\als)$ relativistic corrections with the wave function obtained from the $N=0$ static potential.
{\bf Lower right panel:} As in the lower left panel but with $\nu_r=1$ GeV. 
\label{Fig:hbetabmur}
}   
\end{center}
\end{figure}
We now turn to $n=2$ bottomonium states. Following the discussion in \Sec{sec:RF}, we consider first the renormalon-free combination $M_{h_b}-M_{\eta_b}$. We plot the results of our analysis in \fig{Fig:hbetabmur}. The behavior of the static potential is convergent for this energy combination, though less than in the previous case (higher order corrections could still give non-negligible contributions). This is to be expected since, for the first time, we consider the $n=2$ states where the weak-coupling approximation should be worse (still, $n=2$ P-wave states behave typically better than 2S states). The static potential gives an energy shift $\sim 400$ MeV. 
Setting $\nu_r=1$ GeV does not make the result significantly more scale independent for the static potential (in this respect $\nu_r =0.7$ GeV produces flatter curves). Indeed, the difference between both computations is small for the static potential and also not very different from the strict fixed-order result. The incorporation of the relativistic corrections (and the ultrasoft correction) improves the agreement with experiment. It also increases the dependence on the scale, more for the case with $\nu_r=\infty$ than for the case $\nu_r=1$ GeV (a flatter curve is indeed obtained for a smaller value of $\nu_r=0.7$ GeV, as expected). The case $\nu_r=\infty$ is also quite similar to the strict fixed-order computation. We can estimate the size of the shift of the ${\cal O}(v^4)$ to be of order 200 MeV. The difference with experiment is of order 100 MeV, which is consistent with expected uncertainties. 

\begin{figure}[!htb]
	\begin{center}   
	\includegraphics[width=0.46\textwidth]{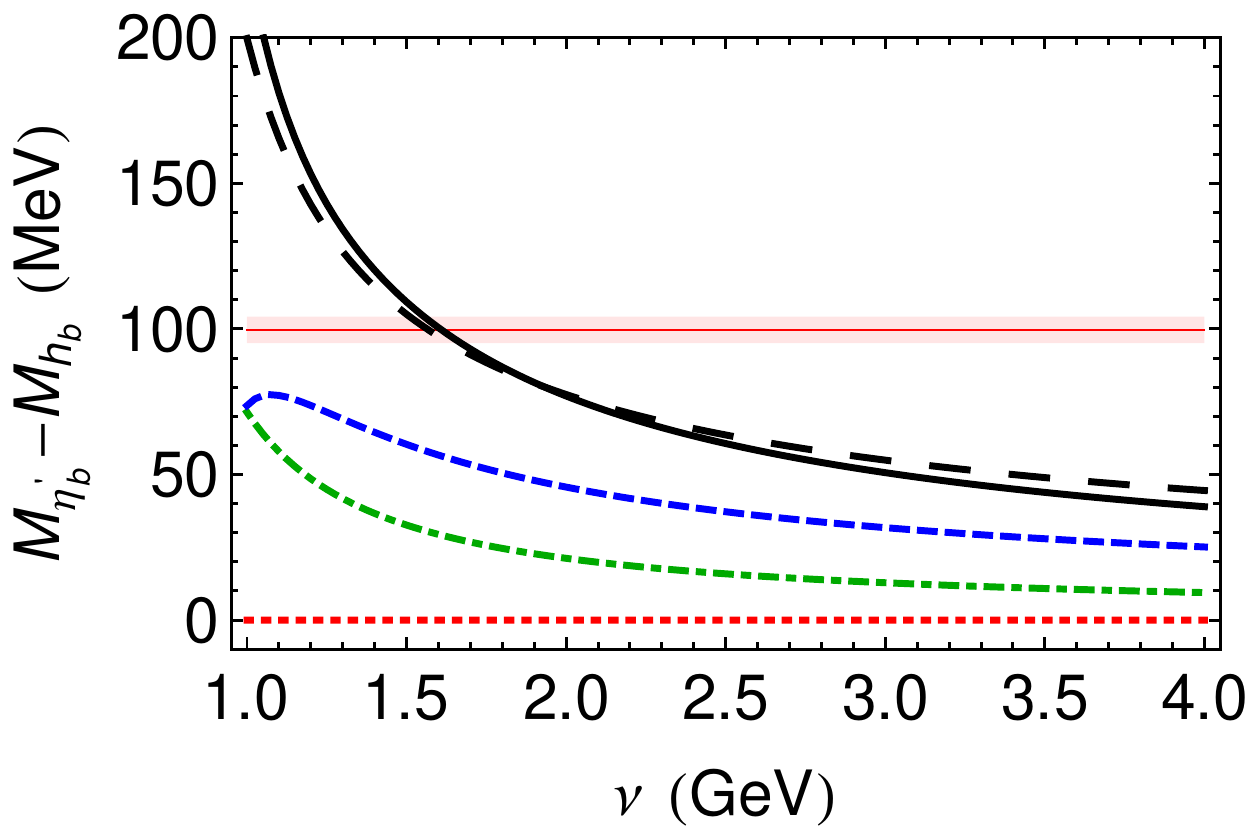}	      
	\includegraphics[width=0.46\textwidth]{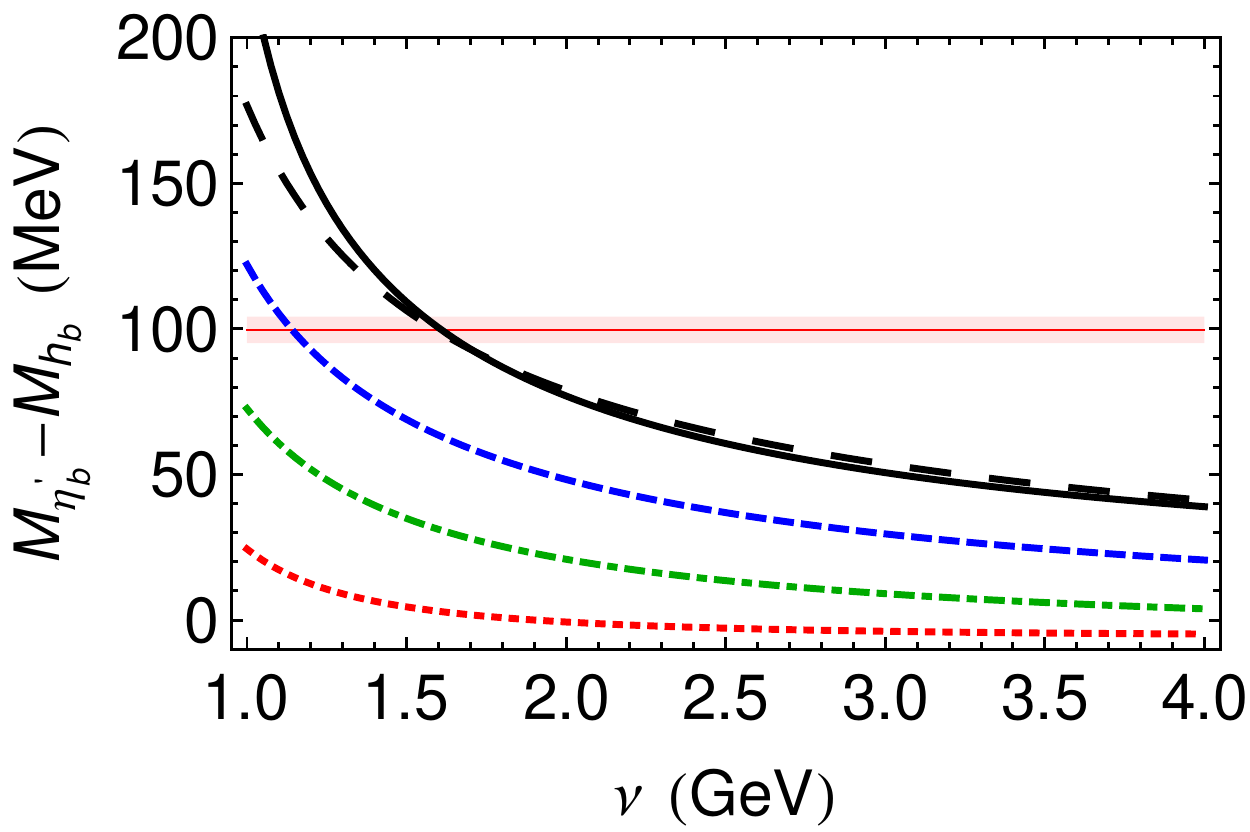}\\
	\includegraphics[width=0.46\textwidth]{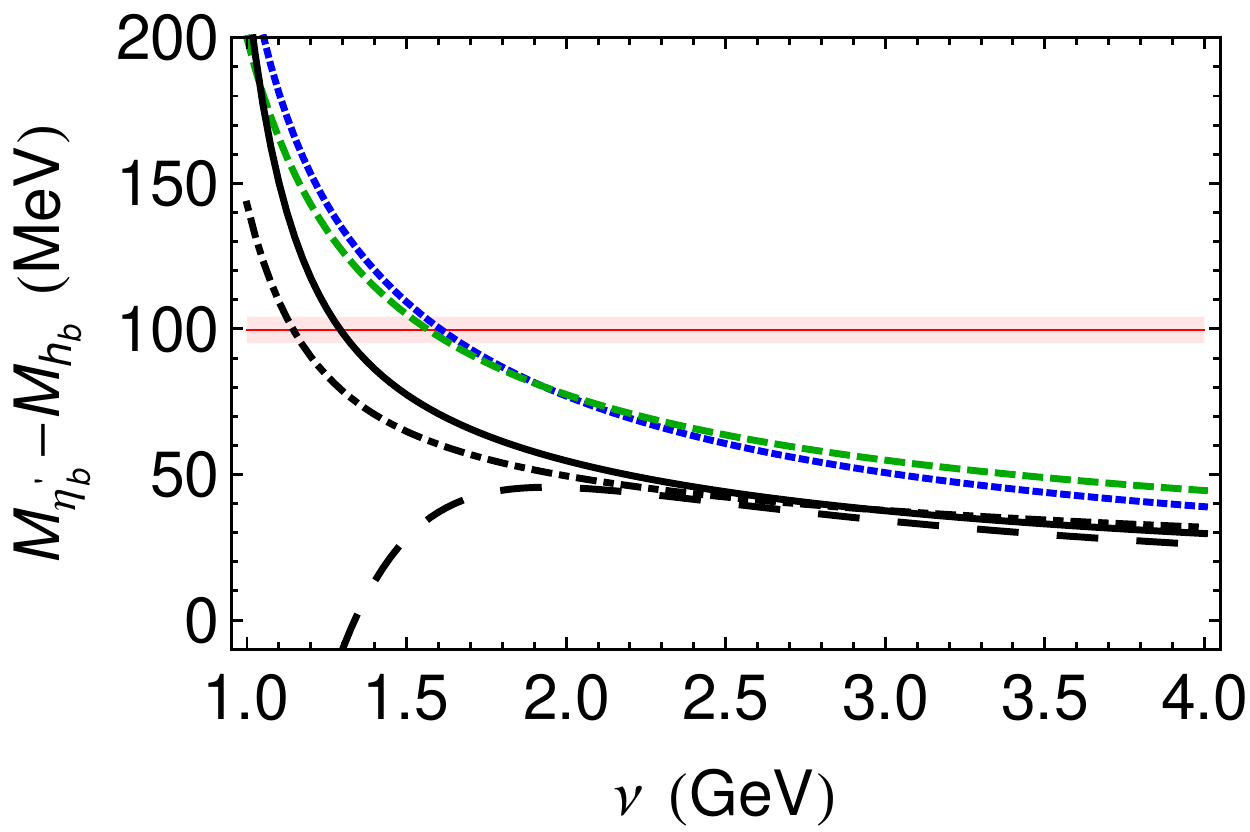}	      
	\includegraphics[width=0.46\textwidth]{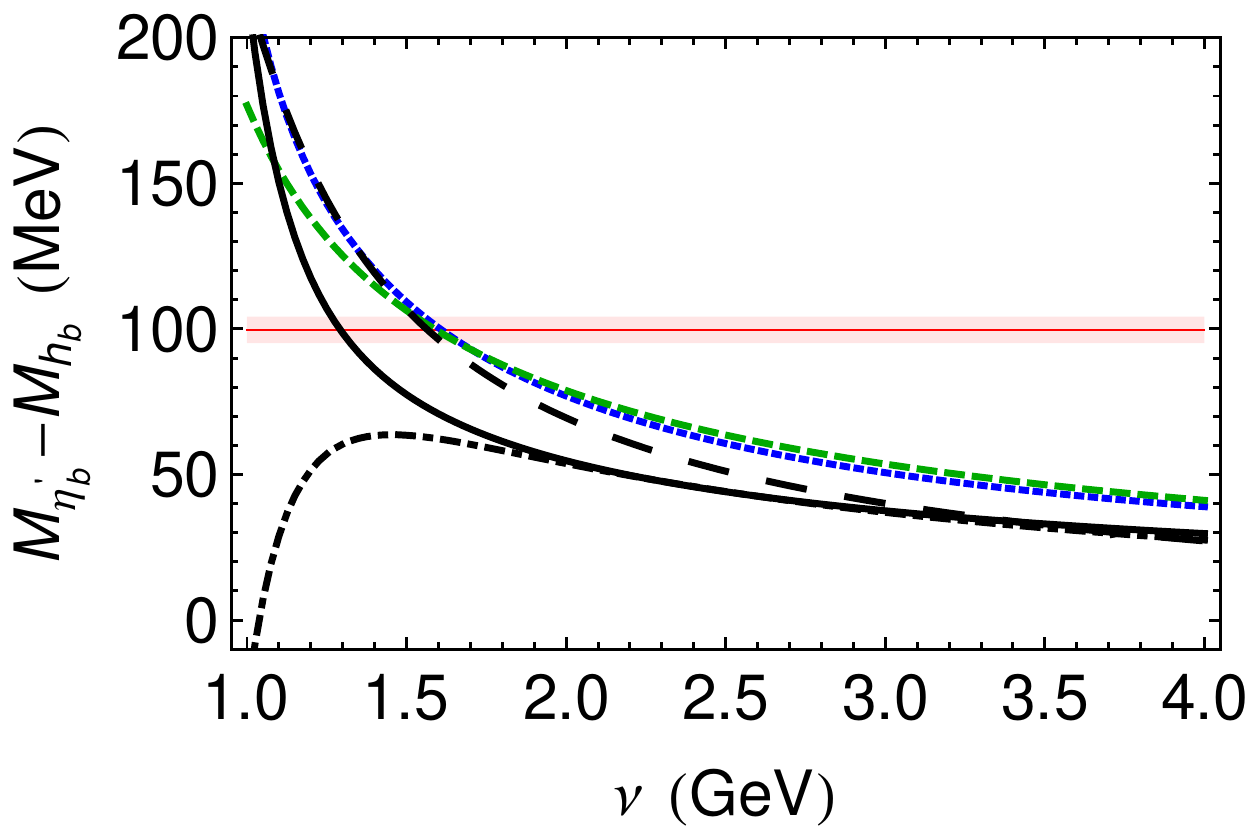}\\
\caption{
Plot of $M_{\eta'_b}-M_{h_b}$ in MeV in RS' scheme for $\nu_f=1$ GeV, $m_{b,\RS'}=4.885$ GeV. Red band is the experimental value.
{\bf Upper left panel:} $\nu_r=\infty$ GeV. Red dotted, green dot-dashed, blue dashed and black long-dashed 
 are the results obtained solving \eq{eq:Schroedinger} with the static potential truncated at N=0,1,2,3, respectively. Solid black line is the $N=3$  previous result in strict fixed-order perturbation theory. 
{\bf Upper right panel:} As in the upper left panel but with $\nu_r=1$ GeV.   
{\bf Lower left panel:} $\nu_r=\infty$ GeV. Blue dotted line corresponds to the solid black line of the above figure, green short-dashed line corresponds to the long-dashed black line of the above figure. The other lines incorporate the relativistic and ultrasoft corrections. Solid black line is the full result in strict fixed-order perturbation theory (solid black line in the right panel in \fig{Fig:SPwavebb}).  The long-dashed black line corresponds to \eq{Mnlfull} evaluating the relativistic corrections with the wave function obtained from the $N=3$ static potential. The dot-dashed black line corresponds to \eq{Mnlfull} evaluating the ${\cal O}(mv^4)$ relativistic corrections with the wave function obtained from the $N=1$ static potential and the ${\cal O}(mv^4\als)$ relativistic corrections with the wave function obtained from the $N=0$ static potential.
{\bf Lower right panel:} As in the lower left panel but with $\nu_r=1$ GeV. \label{Fig:etabhbmur}}   
\end{center}
\end{figure}
Finally, we consider the renormalon-free combination $M_{\eta'_b}-M_{h_b}$. We show the analysis in \fig{Fig:etabhbmur}. 
The behavior of the static potential is not really convergent for this energy combination. This is to be expected since it is unclear whether the weak-coupling expansion is valid for $2S$ states (somewhat it depends on the observable, as we saw in \rcite{Pineda:2013lta}). The scale dependence on $\nu$ of this observable is large, and changing $\nu_r$ does not improve the scale dependence. The difference with the strict fixed-order computation is not large. We cannot really give quantitative estimates. On the other hand, it is fair to say 
that, had we considered $M_{\eta'_b}-M_{\eta_b}$ instead, the general convergence pattern would have been better. The fact that we have considered a more fine-tuned observable has amplified the relative errors.

\section{Conclusions}
\label{concl}
In this paper we have determined  the charm and bottom quark mass fitting the N$^3$LO perturbative expression of the ground state 
(pseudoscalar) energy of the bottomonium, charmonium and $B_c$ systems:
the $\eta_b$, $\eta_c$ and $B_c$ masses, to their experimental values. 
Our result for the $\MSbar$ masses read 
\be
\m_{c}(\m_{c})=1223(33)\; {\rm MeV} \quad {\rm and} \quad 
\m_{b}(\m_{b})=4186(37)\; {\rm MeV}\;. 
\ee
The value of the charm mass is obtained by considering the energy combination: $M_{B_c}-M_{\eta_b}/2$. 

In Figs. \ref{Tabmb} and
\ref{Tabmc} we have compared our numbers with alternative determinations from heavy quarkonium, either from spectroscopy or from sum rules, in the strict weak-coupling approximation. For the case of the bottom quark mass all determinations are consistent with each other. For the case of the charm quark mass our number is perfectly compatible with other determinations from spectroscopy, but there is a small tension with determinations using low $n$ sum rules, where our number is in the low range. In any case, the difference is hardly statistically significant.
 
The consideration of the bottomonium, charmonium and $B_c$ together opens the possibility of using the renormalon-free energy combination: $M_{B_c}-M_{\eta_b}/2-M_{\eta_c}/2$, which is weakly dependent on the heavy quark masses but shows a strong dependence on $\als$. We use this observable to obtain a determination of $\als$: 
\be
\als(M_z)=0.1195(53)\;.
\ee
This number is perfectly compatible with the world average, albeit with larger errors.

We have also explored the applicability of the weak coupling approximation to $n=2$ states. We have limited the analysis to bottomonium.  The N$^3$LO prediction for $M_{h_b}-M_{\eta_b}$ is compatible with experiment albeit the convergence is marginal. For $M_{\eta'_b}-M_{h_b}$ the situation is somewhat worse: the experimental value is reproduced by the N$^3$LO prediction at around $\nu \sim 1.5$ GeV but the convergence is not good and the renormalization scale dependence is rather large.

Finally, we have also studied an alternative computational scheme that reorganizes the perturbative expansion by treating the static potential 
(truncated to order $\als^{N+1}$) exactly. We have observed that the solution coming from the static potential converges (as we increase $N$) for the bottomonium ground state mass for $\nu_f=1$ GeV. Incorporating the resummation of soft logarithms ($\nu=1/r$ for $1/r < \nu_r$) further diminishes the scale dependence. A similar picture holds for the $B_c$ and charmonium if we set $\nu_f=0.7$ GeV (also, but with less convergence, for the P-wave $n=2$ bottomonium state). Nevertheless, the relativistic corrections do not generally show a good behavior. The difference with the corresponding relativistic strict fixed-order computation is large. However, if we turn to renormalon-free energy combinations the situation improves. First, the convergence of the static potential is better, specially for the combinations $M_{B_c}-M_{\eta_b}/2-M_{\eta_c}/2$ and $M_{h_b}-M_{\eta_b}$ (for $M_{\eta'_b}-M_{h_b}$ the improvement is marginal). The relativistic corrections also show a better behavior. The theoretical prediction for $M_{h_b}-M_{\eta_b}$ shows a better agreement with experiment after the inclusion of the relativistic corrections. Nevertheless, this is not really so for $M_{B_c}-M_{\eta_b}/2-M_{\eta_c}/2$, even though the degree of cancellation between very large individual relativistic corrections is still quite remarkable. At this stage we cannot make firm statements about the behavior of the relativistic corrections. We can speculate though that including more terms in the perturbative expansion and/or performing a renormalization group analysis could improve their behavior. 
On top of that, the different behavior between renormalon-sensitive and renormalon-free observables (for which the renormalon cancellation is achieved from the start) may indicate that small inefficiencies in the renormalon cancellation using threshold masses get amplified in this computational scheme: affecting direct determinations of the heavy quarkonium masses, but largely canceling for renormalon-free observables. We hope to come back to all these issues in the near future. 

\begin{acknowledgments}
\noindent
This work was supported
in part by the Spanish grants FPA2014-55613-P, FPA2017-86989-P and SEV-2016-0588. 
J.S. acknowledges the financial support from the European Union's Horizon 2020 research and innovation programme under the Marie Sk\l{}odowska--Curie Grant Agreement No. 665919, and from Spanish MINECO's Juan de la Cierva-Incorporaci\'on programme, Grant Agreement No. IJCI-2016-30028.
\end{acknowledgments}
  
\appendix

\section{The renormalon subtraction}
We perform the renormalon subtraction in the RS and RS' schemes for all the energy levels and for different masses following the procedure in \rcite{Pineda:2001zq,Ayala:2014yxa}. We find that there is a non trivial mass dependence for different masses. The RS(RS') energy levels are: 
\begin{align}
E_{\rm N^iLO,RS(')}=E_{\rm N^iLO}+\delta r_{\rm RS(')}^{(i)}
\end{align}
For different masses, following the notation in \rcite{Ayala:2014yxa} (see also this reference for extra details), we have 
\begin{align}
\delta r_{\rm RS}^{(0)}&=2 \als (\nu)\nu_f N_m \mathcal{C}_0 ,\nn\\
\delta r_{\rm RS}^{(1)}&=\frac{\als^2 (\nu) }{\pi }\nu_f N_m (\beta_0 \mathcal{C}_1+2 z_1 \mathcal{C}_0),\nn\\
\delta r_{\rm RS}^{(2)}&=\frac{\als^3 (\nu) }{2\pi^2} \nu_f N_m \left(-\frac{\pi^2C_F^2 m_r^2 \mathcal{C}_0}{n^2}\left(\frac{1}{m_1^2}+\frac{1}{m_2^2}\right)+\beta_0^2 \mathcal{C}_2+4 \beta_0 z_1 \mathcal{C}_1+4 z_2 \mathcal{C}_0\right),\nn\\
\delta r_{\rm RS}^{(3)}&=-\frac{ \als^4 (\nu) }{4 \pi ^3}\nu_f N_m\left\{\frac{\pi ^2 C_F^2 m_r^2}{n^2}\left(\frac{1}{m_1^2}+\frac{1}{m_2^2}\right)\left(\mathcal{C}_0 \left(2 \beta_0 L_\nu+a_1-2 \beta_0+2 z_1\right)+\beta_0 \mathcal{C}_1\right)\right.\nn\\
&-\beta_0 \left(4 \left(z_1^2+2 z_2\right) \mathcal{C}_1+\beta_0 (\beta_0 \mathcal{C}_3+6 z_1 \mathcal{C}_2)\right)-8 z_3 \mathcal{C}_0\bigg\}
\end{align}
and 
\begin{align}
\delta r_{\rm RS'}^{(0)}&=0,\nn\\
\delta r_{\rm RS'}^{(1)}&=\frac{ \als^2 (\nu)}{\pi } \nu_f N_m \beta_0\mathcal{C}_1,\nn\\
\delta r_{\rm RS'}^{(2)}&=\frac{\als^3 (\nu ) }{2 \pi ^2}( \nu_f N_m \beta_0(\beta_0 \mathcal{C}_2+4 z_1 \mathcal{C}_1)),\nn\\
\delta r_{\rm RS'}^{(3)}&=-\frac{\als^4 (\nu) }{4 \pi ^3} \nu_f N_m \beta_0 \left(\frac{\pi ^2 C_F^2 m_r^2 \mathcal{C}_1}{n^2}\left(\frac{1}{m_1^2}+\frac{1}{m_2^2}\right) -4 \left(z_1^2+2 z_2\right) \mathcal{C}_1-\beta_0 (\beta_0 \mathcal{C}_3+6 z_1 \mathcal{C}_2)\right),
\end{align}
where $L_\nu=\ln(n\nu/(2C_Fm_r\als))+H_{n+l}$, being $H_n$ the $n$th-harmonic number,
\begin{align}
\mathcal{C}_N&=\sum_{n=0}^3 \tilde \chi_n \frac{\Gamma(\bar\nu+N+1-n)}{\Gamma(\bar\nu+1-n)}
\,,
\end{align}
and $\tilde\chi_n$ are the coefficients $\tilde c_n$ in Eqs.~(3.2b) and~(3.2c) in 
\rcite{Ayala:2014yxa}. Note that in that reference the value of $\beta_4$ was not known and an estimate was used. Here we use the value obtained in \cite{Baikov:2016tgj,Herzog:2017ohr}. This indeed changes the value of the RS/RS' masses.

\end{document}